\documentclass[aps,prd,twocolumn,showpacs,groupedaddress]{revtex4}
\usepackage{epsfig}
\usepackage{longtable}
\usepackage{amsfonts}
\newcommand{\Det}{\mbox{det}}
\newcommand{\n}{{\vec n}}
\newcommand{\m}{{\vec m}}
\newcommand{\kk}{{\vec k}}

\newcommand{\oo}{{\omega_\oplus}}
\newcommand{\ot}{\omega_t}
\newcommand{\To}{T_\oplus}
\newcommand{\Oo}{\Omega_\oplus}
\begin{document}

\title{Optical cavity tests of Lorentz invariance for the electron}
\author{Holger M\"{u}ller, Sven Herrmann, Alejandro Saenz, and Achim Peters}
\affiliation{Institut f\"{u}r Physik, Humboldt-Universit\"{a}t zu Berlin, 10117 Berlin, Germany.\\
Tel. +49 (30) 2093-4907, Fax +49 (30) 2093-4718} \email{holger.mueller@physik.hu-berlin.de}
\author{Claus L\"{a}mmerzahl}
\affiliation{ZARM (Center for Applied Space Technology and Microgravity), Universit\"{a}t Bremen,
Am Fallturm, 28359 Bremen, Germany}


\begin{abstract}
A hypothetical violation of Lorentz invariance in the electrons' equation of motion (expressed
within the Lorentz-violating extension of the standard model) leads to a change of the geometry of
crystals and thus shifts the resonance frequency of an electromagnetic cavity. This allows
experimental tests of Lorentz invariance of the electron sector of the standard model. The
material dependence of the effect allows to separate it from an additional shift caused by Lorentz
violation in electrodynamics, and to place independent limits on both effects. From present
experiments, upper limits on Lorentz violation in the electrons' kinetic energy term are deduced.
\end{abstract}


\maketitle


\section{Introduction}
Special relativity and the principle of Lorentz invariance describe how the concepts of space and
time have to be applied when describing physical phenomena in flat space--time. Improving the
accuracy of the experimental verification of these fundamental concepts is of great interest, also
because a violation of Lorentz invariance is be a feature of many current models for a quantum
theory of gravity, e.g., string theory \cite{kos89,ellis}, loop gravity \cite{gambini,alfaro}, and
non--commutative geometry \cite{ncg}. Such a violation of Lorentz invariance is described in the
general standard model extension (SME) developed by Colladay and Kosteleck\'{y} \cite{KostSME}.
According to it, Lorentz violating terms might enter the equations of motion of bosons and
fermions. At first sight, the quantum gravity induced corrections and effects are of order
$E/E_{\rm  QG} \sim 10^{-28}$ where $E$ is the energy scale of the experiment (which in ordinary
optical experiments is of the order 1 eV) and $E_{\rm QG} \sim E_{\rm Planck}$ is of the order of
the Planck energy. Therefore these effects seem to be far from being observable in laboratory
experiments. However, as it occurs e.g. in scenarios leading to a modification of the Newton
potential at small distances, some mechanism may apply which effectively leads to much larger
effects in the laboratory. It is thus interesting to find experimental configurations in the
laboratory that can place strong upper limits on as many of these terms as possible.

Experiments on Lorentz symmetry that study light propagation have a long and fascinating history,
starting from the original interferometer experiments of Michelson \cite{Michelson81} in Potsdam
and Michelson and Morley \cite{MM} in Cleveland. Modern versions of these experiments
\cite{BrilletHall,HilsHall,Braxmaier,Wolf,Lipa,MuellerASTROD,MuellerMM} replace the interferometer
by a measurement of the resonance frequency $\nu = m c/(2L)$ of an electromagnetic
(Fabry-P\'{e}rot) cavity, that is given by the velocity of light $c$ along the cavity axis, the
cavity length $L$, and a constant mode number $m$. Lorentz violation causing a shift of $c$ or $L$
connected to a rotation or boost of the cavity frame of reference can thus be detected through the
corresponding shift of the resonance frequency. From such experiments, upper limits on a tensor
$(k_F)_{\kappa \lambda \mu \nu}$ have been found, that encodes Lorentz violation in the photonic
sector of the SME \cite{MuellerMM,Wolf,Lipa,MMlang,KosteleckyMewesPRL,KosteleckyMewesPRD}. These
experiments are mainly based on the shift of $c$ connected to non-zero values of $(k_F)_{\kappa
\lambda \mu \nu}$, as an additional change of $L$ caused by $(k_F)_{\kappa \lambda \mu \nu}$ and a
corresponding orientation dependent modification of the Coulomb potential is negligible for most
cavity materials \cite{TTSME}.

In this work, we treat the effect of Lorentz violation within the fermionic sector of the standard
model extension in cavity experiments. A modified kinetic energy term entering the
non-relativistic Schr\"{o}dinger hamiltonian of the free electron $(p^2+2E'_{jk} p_j p_k) /(2m)$
leads to a change of the geometry of crystals, and thus a change $\delta_{e-} \nu$ of the
resonance frequency of a cavity made from this crystal. Here, $p_j$ is the 3-momentum, $m$ the
electron mass, and $E'_{jk}=-c_{jk}-c_{00} \delta_{jk}$ a dimensionless $3\times 3$ matrix given
by a tensor $c_{\mu \nu}$ of the SME. Thus, the total shift of the resonance frequency $\delta
\nu=\delta_{e-} \nu_{res} + \delta_{\rm EM}\nu_{res}$, where $\delta_{\rm EM} \nu$ denotes the
shift due to Lorentz violation in the electromagnetic sector. Since $\delta_{e-} \nu$ depends on
the cavity material, it can be be distinguished from Lorentz violation in electrodynamics by
comparing cavities made from different materials. Experiments using suitable configurations of
cavities can place separate upper limits on the components of $c_{\mu \nu}$ and $(k_F)_{\kappa
\lambda \mu \nu}$. Using data available from past experiments, we deduce approximate bounds on
some combinations of components of $c_{\mu \nu}$ at the $10^{-14}$ level. From future cavity
experiments, Earth and space-based, that are projected as tests of Lorentz violation in
electrodynamics \cite{OPTIS,SUMO,BluhmSpaceTests}, bounds at a level down to $10^{-18}$ can be
expected.

In Sec. \ref{smesection}, we present the non--relativistic hamiltonian of the free electron within
the SME. Since violations of Lorentz invariance are certainly small, it is sufficient to work to
first order in the Lorentz--violating modifications throughout. The crystal adjusts its geometry
such as to minimize the total energy $\left< \delta h \right>+U_{\rm elast}$, where $\left< \delta
h \right>$ is the expectation value of the Lorentz--violating part of the hamiltonian and $U_{\rm
elast}$ is the elastic energy associated with distortion of the crystal. It is calculated in Sec.
\ref{deltahmatrixelements}. The resulting geometry change is calculated in Sec.
\ref{lengthchange}. A fairly detailed model for the crystal allows us to obtain specific results
for practical materials, including quartz and sapphire. In Sec. \ref{experiments}, we discuss
experimental configurations and obtain bounds on Lorentz violation in the electrons' equation of
motion from present experiments. In appendix \ref{spinpolarization}, we discuss the hypothetical
case of a cavity made from a spin-polarized solid, which allows to place experimental limits on an
additional, spin dependent term from the SME, at least in principle. In appendix
\ref{notationconvention}, we summarize some conventions made in elasticity theory that are needed
for our calculations, and in appendix \ref{sigcomp}, we give in detail the Fourier components of
the signal for Lorentz violation in laboratory experiments on Earth.


\section{Standard model extension}\label{smesection}

\subsection{Model}

The SME starts from a Lagrangian formulation of the standard model, adding all possible observer
Lorentz scalars that can be formed from the known particles and Lorentz tensors. Taken from the
full SME that contains all known particles, the Lagrangian involving the Dirac fields $\psi^e$ of
the electron and $\psi^p$ of the proton and the electromagnetic field $F^{\mu\nu}$ can be written
as (in this section, we use units with $\hbar=c=1$; the greek indices take the values $0,1,2,3$)
\cite{KostSME,Kostelecky99}
\begin{eqnarray}\label{lagrangian}
{\mathcal L} & = & \frac i2 \bar \psi^e \Gamma_\nu^{e} D^\nu \psi^e -\frac 12 \bar \psi^e M^e
\psi^e + \frac i2 \bar \psi^p \Gamma_\nu^p D^\nu \psi^p \nonumber \\ & &  -\frac 12 \bar \psi^p
M^p \psi^p + \mbox{h.c}  -\frac14 F^{\mu\nu}F_{\mu\nu} \\ & &
-\frac14(k_F)_{\kappa\lambda\mu\nu}F^{\kappa\lambda}F^{\mu\nu}+\frac12(k_{AF})^\kappa\epsilon_{\kappa
\lambda\mu\nu}A^\lambda F^{\mu\nu} \, , \nonumber
\end{eqnarray}
where h.c. denotes the hermitian conjugate of the previous terms, and $A^\lambda$ is the vector
potential. The symbols $\Gamma_\nu^{e,p}$ and $M^{e,p}$ are given by
\begin{eqnarray}\label{lagrangianparameters}
\Gamma_\nu & = & \gamma_\nu + c_{\mu \nu} \gamma^\mu + d_{\mu \nu} \gamma_5 \gamma^\mu+ e_\nu
+if_\nu \gamma_5 + \frac 12 g_{\lambda \mu \nu} \sigma^{\lambda \mu} \, , \nonumber \\ M & = & m +
a_\mu \gamma^\mu + b_\mu \gamma_5 \gamma^\mu + \frac 12 H_{\mu \nu} \sigma^{\mu \nu} \, ,
\end{eqnarray}
where the superscripts $e$ and $p$ are to be added to the symbols $a_\mu, b_\mu, c_{\mu \nu},
d_{\mu\nu}, e_\mu, f_\mu, g_{\lambda \mu \nu}$, and $H_{\mu \nu}$ that represent tensors encoding
Lorentz violation for the fermions. $m^e$ and $m^p$ are the electron and the proton mass,
$\gamma_\nu, \gamma_5$ and $\sigma^{\mu \nu}$ are the conventional Dirac matrices, and $D^\nu$ is
the covariant derivative. In this work, we deal mostly with electrons and add a superscript to
denote parameters for particles other than the electron, e.g., $c_{\mu \nu}$ is a parameter for
the electron and $c^p_{\mu \nu}$ the corresponding parameter for the proton. The tensors entering
$M$ have the dimension mass, the others are dimensionless. $H_{\mu \nu}$ is antisymmetric;
$g_{\lambda \mu \nu}$ is antisymmetric in its first two indices. $c_{\mu \nu}$ and $d_{\mu \nu}$
are traceless. Gauge invariance and renormalizability excludes $e_\nu, f_\nu$, and $g_{\lambda \mu
\nu}$, so these are either zero or suppressed relative to the others \cite{Kostelecky99}.

Lorentz violation for the photons is encoded in the tensors $(k_{AF})^\kappa$ and
$(k_F)_{\kappa\lambda\mu\nu}$. The four degrees of freedom contained in $(k_{AF})^\kappa$ are
constrained strongly in measurements of cosmological birefringence
\cite{KosteleckyMewesPRL,KosteleckyMewesPRD} and are neglected in what follows. 10 of the 19
degrees of freedom of $(k_F)_{\kappa\lambda\mu\nu}$ are constrained by astrophysical observations
\cite{KosteleckyMewesPRL,KosteleckyMewesPRD}, the other 9 can be measured in cavity experiments
\cite{KosteleckyMewesPRL,KosteleckyMewesPRD,Lipa,MuellerMM,MMlang}.

\subsection{Modified non-relativistic hamiltonian}

\subsubsection{Free electron}
The non-relativistic Schr\"{o}dinger hamiltonian $h=\hat{h}+\delta h$ of a single free electron
within the SME derived from this Lagrangian (using Foldy-Wouthuysen methods,
\cite{Kosteleckypreprint}) is the sum of the usual free-particle Hamiltonian $\hat h$ and a
Lorentz-violating term \cite{Kosteleckypreprint,Kostelecky99}
\begin{eqnarray}\label{deltah}
\delta h & = & m A' + m B'_j\sigma^j+C'_j p_j + D'_{jk} p_j \sigma^k \nonumber \\ & & +E'_{jk}
\frac{p_j p_k}{m} + F'_{jkl} \frac{p_j p_k}{m}\sigma^l
\end{eqnarray}
with the components of the 3-momentum $p_j$ and of the Pauli matrices $\sigma^j$. The latin
indices take the values $1,2,3$. (We denote both 3-vectors such as $x_j$ and reciprocal 3-vectors
such as $p_j$ by subscript.) A Hamiltonian of this form has also been derived in
\cite{LaemmerzahlCQG}. The constant term $mA'$ has no physical consequences and is included for
completeness. The term proportional to $C'_j$ can be eliminated by choosing coordinates such that
the systems centre of mass is at rest \cite{LaemmerzahlCQG}. The dimensionless coefficients
$A',B'_j,C'_j,D'_{jk},E'_{jk}$ and $F'_{jkl}$ can be expressed in terms of the quantities entering
the Lagrangian \cite{Kosteleckypreprint,Kostelecky99}:
\begin{eqnarray}
A' &=& \frac 1m a_0 -c_{00} - e_0 \, , \\ B'_j & = & -\frac{b_j}{m} + d_{j0} - \frac12
\varepsilon_{jkl}g_{kl0} + \frac{1}{2m} \varepsilon_{jkl}H_{kl} \, , \\ C'_j & = & -\frac{a_j}{m}+
(c_{0j}+c_{j0}) + e_j \, , \\ D'_{jk} & = &  \frac{b_0}{m} \delta_{jk} - (d_{kj}
+d_{00}\delta_{jk}) \nonumber \\ & & - \varepsilon_{klm}(\frac 12 g_{mlj}+g_{m00}\delta_{jl})
-\frac 1m \varepsilon_{jkl} H_{l0} \, , \\ E'_{jk} & = &-c_{jk}-\frac{1}{2}c_{00}\delta_{jk} \, ,
\label{hamiltonianparameters} \\ F'_{jkl} & = &\bigg[(d_{0j}+d_{j0}) -\frac12\bigg(
\frac{b_j}{m}+d_{j0}+\frac12 \varepsilon_{jmn}g_{mn0} \\ & & +\frac{1}{2m}\varepsilon_{jmn}H_{mn}
\bigg)\bigg] \delta_{kl}+\frac 12\left(\frac{b_l}{m}+\frac
12\varepsilon_{lmn}g_{mn0}\right)\delta_{jk} \nonumber \\ & & -\varepsilon_{jlm} (g_{m0k}+g_{mk0})
\, . \nonumber
\end{eqnarray}

\subsubsection{Interaction terms}
In addition to $\delta h$, the Hamiltonian arising from the Lagrangian Eq. (\ref{lagrangian}) also
involves modified interaction terms proportional to combinations of $a_\mu, b_\mu, c_{\mu \nu},
d_{\mu\nu}, e_\mu, f_\mu, g_{\lambda \mu \nu}$, and $H_{\mu \nu}$. For the non-relativistic
electrons in solids, however, these are suppressed by a factor given by $\alpha$, the
fine-structure constant, relative to the modifications of the free-particle Hamiltoninian, Eq.
(\ref{deltah}) \cite{Kostelecky99}. This is basically because the typical energy scale for such
electrons is the Rydberg energy $m\alpha^2/2$. We can therefore neglect the modified interaction
terms.

\subsection{Coordinate and field redefinitions}

Some of the parameters contained in either the photon, electron, or proton sectors of the
Lagrangian Eq. (\ref{lagrangian}) can be absorbed into the other sectors by coordinate and field
redefinitions without loss of generality. Thus, not all of the coefficients in the Lagrangian have
separate physical meanings. Loosely speaking, in experiments where one compares the sectors against
each other only differential effects are meaningful.

For example, in a hypothetical world containing only photons and electrons, the nine components of
$(k_F)_{\kappa\lambda\mu\nu}$ not constrained by astrophysical experiments could be absorbed into
the nine symmetric components of $c_{\mu\nu}^e$
\cite{KostSME,KosteleckyMewesPRL,KosteleckyMewesPRD,KosteleckyPickering}. By definition, either
the photon or the electron sector could be taken as conventional with all the Lorentz violation in
the other sector. For example, for tests of Lorentz violation for the photon
\cite{KosteleckyMewesPRL,KosteleckyMewesPRD,Lipa,MuellerMM,MMlang}, one implicitly
assumes a conventional electron sector.

The presence of protons (and neutrons) in the solid changes this picture. We can still assume that
one of the sectors is conventional, but then in general the other sectors are Lorentz-violating.
Choosing a conventional proton sector allows us to disregard the proton terms. It also fixes the
definition of coordinates and fields so that the components of $c_{\mu\nu}$ cannot be absorbed into
$(k_F)_{\kappa\lambda\mu\nu}$ in general, i.e. they acquire separate physical meanings.

To illustrate this, it suffices to consider an extension of the toy version of the SME introduced
in \cite{KosteleckyMewesPRD}. Its Lagrangian describes electrons and protons as scalar fields
$\phi^e$ and $\phi^p$, neglecting spin effects:
\begin{eqnarray}
\mathcal L^\phi &= &(\eta_{\mu\nu}+k_{\mu\nu})(D_\mu\phi^e)^\dag
D_\nu\phi^e-m^2(\phi^e)^\dag\phi^e \nonumber \\ & & +(D_\mu\phi^p)^\dag
D^\mu\phi^p-(m^p)^2(\phi^p)^\dag\phi^p \\ & &
-\frac14F^{\mu\nu}F_{\mu\nu}-\frac14(k_F)_{\kappa\lambda\mu\nu}F^{\kappa\lambda}F^{\mu\nu} \, .
\nonumber
\end{eqnarray}
It has a conventional proton sector and non-conventional electron and photon sectors. Lorentz
violation for the electron is given by the coefficient $k_{\mu\nu}$. As usual, the covariant
derivative is given by $D_\mu \phi^{e,p}=(\partial_\mu +iq^{e,p} A_\mu)\phi^{e,p}$, where
$q^{e,p}$ is the particle's electric charge. If one identifies $k_{\mu\nu}=c_{\mu\nu}^e$, this
Lagrangian leads to the modified Hamiltonian Eq. (\ref{deltah}) if only the $c_{\mu\nu}$ are
nonzero.

For simplicity, consider the special case of only one nonzero component $k_{00}=k^2-1$, where $k$
deviates slightly from unity. The lagrangian takes the form
\cite{KosteleckyMewesPRD}
\begin{eqnarray}\label{toylagrangian}
\mathcal L^\phi & = & (D_\mu\phi^e)^\dag D_\nu\phi^e +(k^2-1)|D_0\phi^e|^2-m^2_e(\phi^e)^\dag
\phi^e \nonumber
\\ & &  +(D_\mu\phi^p)^\dag D^\mu\phi^p-(m^p)^2(\phi^p)^\dag\phi^p \\ & &
-\frac14F^{\mu\nu}F_{\mu\nu}-\frac14(k_F)_{\kappa\lambda\mu\nu}F^{\kappa\lambda}F^{\mu\nu} \, .
\nonumber
\end{eqnarray}
By coordinate transformations $t\rightarrow tk$, $\vec x \rightarrow \vec x$, the field
redefinition $A_0 \rightarrow A_0$, $A_i \rightarrow A_i k$, and rescaling the electric charge $q
\rightarrow q/k$, one obtains the Lagrangian
\begin{eqnarray}\label{redeftoy}
\mathcal L^\phi &= & (D_\mu\phi^e)^\dag D_\nu\phi^e -m^2(\phi^e)^\dag \phi^e +(D_\mu\phi^p)^\dag
D^\mu\phi^p\nonumber \\ & & +(k^{-2}-1)|D_0\phi^p|^2-(m^p)^2(\phi^p)^\dag\phi^p \\ & &
-\frac14F^{\mu\nu}F_{\mu\nu}-\frac14(k_F)_{\kappa\lambda\mu\nu}F^{\kappa\lambda}F^{\mu\nu}
+\frac{1-k^2}{2}B^2 \, , \nonumber
\end{eqnarray}
where $B$ is the magnetic field. Thus, the Lorentz violation in the electron sector has been moved
to the proton and photon sectors. (If $(k_F)_{\kappa\lambda\mu\nu}=0$, the parameter $1/k$ can be
interpreted as a modified velocity of light \cite{KosteleckyMewesPRD}.) However, it is in general
not possible to eliminate Lorentz violation in more than one sector at the same time.

Cavity experiments compare the velocity of a light wave to a length defined by a crystal. In the
light of Eq. (\ref{toylagrangian}), the Lorentz violation for the electron acts via the term
$(k^2-1)|D_0\phi^e|^2$. With a time-independent Coulomb potential $A_0 =$ const., this contributes
a term
\begin{equation}\label{toymodcoulomb}
-2im(k^2-1)[\phi_{,0}+q\Re(A_0)\phi]
\end{equation}
to the equation of motion for $\Phi^e=e^{-imx_0} \phi^e$ in the non-relativistic limit (obtained,
in the usual way, by the Euler-Lagrange equations and setting to zero terms of order $|A_0|^2$ and
$m^0$). $\Re$ denotes the real part. The second term modifies the coupling of the electron to the
Coulomb potential, causing a geometry change of the crystal. Thus, a combination of $k^2-1$ and the
modified velocity of light given by $(k_F)_{\kappa\lambda\mu\nu}$ is measured in the experiment.

In the alternative description by Eq. (\ref{redeftoy}), the same Lorentz violation acts via a term
analogous to Eq. (\ref{toymodcoulomb}) in the equation for the proton, i.e., a rescaled coupling of
the proton to the electric field, and a modified velocity of light given by
$(k_F)_{\kappa\lambda\mu\nu}$ and $k^2-1$. Physically, both pictures are equivalent.

Here, we considered only a single parameter analogous to $c_{00}$, that causes a scaling of the
solid that is rotation-invariant. Thus, it cannot be measured in usual cavity experiments, that
search for a modulation of the effect connected to a rotation of the cavity in space. However, the
tensors $c_{\mu\nu}$ emerging from our special case via the three Lorentz boosts can --- i.e., at
least three out of nine degrees of freedom contained in the symmetric part of $c_{\mu\nu}$. It is
not impossible that, by coordinate and field redefinitions, some of the other parameters can be
absorbed into quantities that have no measurable effect. However, as we have shown, at least 12 out
of 18 parameters from the photon and electron sector (restricting the electron sector to those
parameters that are not constrained by cosmological experiments) would be separately measurable,
that can e.g. be chosen as three $c_{\mu\nu}$ and nine $(k_F)_{\kappa\lambda\mu\nu}$.

In what follows, we adopt a conventional proton sector, with all the Lorentz violation in the
electron and photon sectors. One could possibly extract the measurable quantities from $c_{\mu\nu}$
and consider only those in what follows; however, thereby one would single out a preferred frame in
which the measurable quantities are defined, and loose the covariance under observer Lorentz
transformations which otherwise holds in the SME. Therefore, we choose not to do so and treat all
elements of $c_{\mu\nu}$ as independent.

\subsection{Previous experimental limits on electron parameters}

It is convenient to express limits on the coefficients within a sun--centered celestial equatorial
reference frame as defined in \cite{Kostelecky99}. The components of quantities given in that
frame are denoted by capital indices. Limits for many particles, including muons, protons, and
neutrons, have been studied, see \cite{Kostelecky99,KostPRLmu2000,BluhmSpaceTests} and references
therein. For the electron, the limits given below have been found. However, to our knowledge there
are no experimental limits on $E'_{jk}$ and on many components of $F'_{jkl}$ for the electron.

From clock comparison experiments \cite{Kostelecky99}, a limit on $B_J \lesssim 10^{-24}$ ($m B_J$
is denoted $\tilde b_J$ in\cite{Kostelecky99}) is obtained. Furthermore, for the linear
combinations
\begin{eqnarray}\label{tildeparameters}
\tilde d_J & = & m(d_{0J}+d_{J0})-\frac 12 \left(md_{J0}+\frac 12 \varepsilon_{JKL} H_{KL}
\right)\, , \nonumber
\\ \tilde g_{D,J} & = & m \varepsilon_{JKL} (g_{K0L}+\frac 12 g_{KL0})
 -b_J\, ,
\end{eqnarray}
$\tilde d_J / m\lesssim 10^{-19}$ and $\tilde g_{D,J}/m \lesssim  10^{-19}$. These are
order-of-magnitude limits, since some assumptions are needed to extract them from the measurements
\cite{Kostelecky99}.

An experiment using spin polarized solids yielded $|B'_Z| \simeq (2.7 \pm 1.6) \times 10^{-25}$
\cite{Bluhm00,Adelberger}; in a similar experiment \cite{LiShing}, $((B'_X)^2+(B'_Y)^2)^{1/2} \leq
6.0 \times 10^{-26}$ and $|B'_Z| \leq 1.4 \times 10^{-25}$ have been found.

Hydrogen spectroscopy could prospectively limit linear combinations of $B'_J, B_J^{'p}, d_{J0},
d_{J0}^p, H_{JK}$, and $H^p_{JK}$ (where the superscripts $p$ denotes parameters for the proton)
to about $10^{-27}$\,GeV \cite{Bluhm99}. Comparing the frequencies of hydrogen masers,
\cite{Phillips} find $| B_J^{'p} m^p + B'_J m| \lesssim 2 \times 10^{-28}$\,GeV ($m$ and $m^ p$ are
the electron and proton mass, respectively).

The potential for further tests of Lorentz invariance in space is discussed in
\cite{Bluhmhepph0306190}; for the electron, limits on several parameter combinations are expected
using $^{133}$Cs and $^{87}$Rb clocks. However, these tests allow no limits on the components of
$E'_{jk}$.


\section{Influence of $\delta h$ on solids}\label{deltahmatrixelements}

To first order in the changes, the influence of Lorentz violation in the electron's equation of
motion on the properties of a crystal is induced by the expectation values of the
Lorentz-violating contribution to the hamiltonian, that is calculated in this section. In Sec.
\ref{wavefunctionsect}, we present our ansatz for the electron wave function; the expectation
value $\left<\delta h\right>$ is then calculated in Sec. \ref{matrixelements}.

We denote by $(x_a)_i$, $(p_a)_i$, and $(\sigma_a)^i$ the spatial, 3-momentum, and Pauli matrices
for the $a$\,th particle. The non-relativistic single-particle hamiltonian for the $a$\,th
particle is denoted $h_a= \hat h_a+\delta h_a$. The hamiltonian of the solid
\begin{equation}
h_{\rm all}= \sum_a \left[\hat h_a+ \delta h_a \right]+ \frac 12 \sum_{a \neq b} \left[\hat h_{
a,b}+\delta h_{a,b} \right]
\end{equation}
is the sum of $\hat h_a+\delta h_a$ over all particles, plus the sum of the interaction terms
$\hat h_{a,b}$ over all pairs, and over $\delta h_{a,b}$, a possible Lorentz-violating correction
to it. (The factor $\frac 12$ corrects for the double-counting of pairs.) The Lorentz violating
terms are contained in $\delta h_a$ and $\delta h_{a,b}$. To first order in the changes, the
resulting modifications of the properties of the solid are the sum of the modifications arising
form the individual terms.

The interactions in a solid are electromagnetic. The geometry change of crystals as a consequence
of the modification to the interaction term form the photonic sector of the SME
\cite{KosteleckyMewesPRL,KosteleckyMewesPRD} has been treated for ionic crystals in \cite{TTSME}.
We will not consider this term further here. In this work, we deal with the modifications due to
the Lorentz violation in the electrons' equation of motion, $\sum_a \delta h_a$.



\subsection{Wave function ansatz for the solid}\label{wavefunctionsect}

According to the Bloch theorem (\cite{Ashcroft}, pp. 133-141), the single-electron wave function
$\psi_a$ for the $a$-th electron ($a=1, \ldots, N)$ of a solid can be written as the product of a
plane wave $\exp\{\frac i \hbar \vec q_a \vec x_a \}$ (where $\vec q_a$ is the quasi-momentum of
the $a$\,th electron) and a function $u_{\vec q} (\vec r)$ with the period of the lattice.
$u_{\vec q_a}(\vec r)$ depends on $\vec q_a$, and thus on the electron number $a$. To make a
Fourier expansion of $u_{\vec q_a}(\vec r)$, we note that, if $k_{ji}$ is the $3 \times 3$ matrix
containing the primitive reciprocal lattice vectors $\kk_i$, any reciprocal lattice vector can be
expressed as a linear combination $n_i k_{ji}$ with some coefficients $n_i \in \mathbb Z$
(\cite{Ashcroft}, pp. 86-87). Therefore,
\begin{equation}\label{wavefunction}
\psi_a=\frac{1}{\sqrt{V}}\exp\left\{ \frac i \hbar \vec q_a \vec x_a \right\} \sum_{\vec n}
(c_{\vec n})_a \exp\{-i n_j k_{ij} (x_a)_i \}\, ,
\end{equation}
where $V$ is the volume of the (macroscopic) solid considered. The $(c_{\vec n})_a$ are the
Fourier coefficients of $u_{\vec q_a}(\vec r)$; they depend only on $n_j k_{ij} -(q_a)_i$, i.e.,
they can be expressed as $c_{n_j k_{ij} -(q_a)_i}$. The $\n$ summation is carried out over
$\mathbb Z^3$. The $(c_n)_a$ satisfy
\begin{equation}\label{cnnorm}
\sum_{\vec n} |(c_{\vec n})_a|^2=1
\end{equation}
because of normalization $\left<(\psi)_a|(\psi)_a\right>=1$. If we assume that the crystal has
inversion symmetry, the origin of the coordinate system may be chosen such that (\cite{Ashcroft},
p.137)
\begin{equation}\label{cminusn}
(c_{-\n})_a=(c_\n^*)_a \, .
\end{equation}
The star denotes complex conjugation.

The normalized antisymmetric $N$-electron state $\Psi^N$ ($N$ is the total number of electrons)
can be constructed from the $N \times N$ matrix
\begin{equation}\label{psimatrix}
(\psi) := \left( \begin{array}{ccc}
  \psi_1(\vec x_1) & \ldots & \psi_1(\vec x_N) \\
  \vdots & \ddots & \vdots \\
  \psi_N(\vec x_1) & \ldots & \psi_N(\vec x_N) \\
 \end{array}\right)
\end{equation}
as the Slater determinant \cite{Landau} $\Psi^N=\frac{1}{\sqrt{N!}} \Det(\psi)$ [det denotes the
determinant of a square matrix].


\subsection{Calculation of matrix elements}\label{matrixelements}

\subsubsection{Specifications}

For a bound system in its rest frame, the expectation value of the particle momenta vanishes
\begin{equation}\label{momentumiszero}
\sum_a \left<(p_a)_i \right>=0\, .
\end{equation}
We also assume no spin polarization, i.e., the sum of the spin expectation values
\begin{equation}
S^l \equiv \frac 1N \sum_a \left< (\sigma_a)^l \right>
\end{equation}
vanishes. The case of spin polarization will be considered in the appendix. Furthermore, we assume
a vanishing sum of the helicities of the electrons,
\begin{equation}\label{helicity}
\sum_a \left< (p_a)_j (\sigma_a)^k \right>=0
\end{equation}
and
\begin{equation}
\sum_a \left< (p_a)_i (p_a)_j (\sigma_a)^k \right>=0 \, .
\end{equation}
Although situations could be imagined which violate the last two relations in spite of $S^l= 0$,
this can be considered highly unrealistic. Therefore, the expectation values of the terms
proportional to $C_j$, $D_{jk}$, and $F'_{jkl}$ from the hamiltonian, Eq. (\ref{deltah}), vanish.
Disregarding the constant term proportional to $A$,
\begin{eqnarray}\label{deltahexpect}
\left<(\delta h)\right> & = & \frac{1}{m} E'_{ij} \sum_{a=1}^N \left<(p_a)_i(p_a)_j\right>
\nonumber \, .
\end{eqnarray}

\subsubsection{Calculation of $\delta h$}

We now calculate the matrix element $\left<(p_1)_i(p_1)_j\right>$ for the first electron. Since it
turns out to be independent from the electrons number (a consequence of the antisymmetry of the
$N$-electron state $\Psi^N$), the sum of the matrix elements for all electrons can then be
obtained by multiplying $\left<(p_1)_i(p_1)_j\right>$ with the total number of electrons $N$. We
have
\begin{equation}
\left<(p_1)_i(p_1)_j\right>=\int_V  \Psi^{N*} \frac{-\hbar^2\partial^2}{\partial (x_1)_i (x_1)_j}
\Psi^N d^3 x_1 \ldots d^3x_N \, .
\end{equation}
The integrations are carried out over the volume $V$ of the solid. $\Psi^N$ and $\Psi^{N*}$ are
given by Slater determinants. Evaluation of the matrix element starts by an expansion of these
determinants with respect to the first column,
\begin{equation}
\Det (\psi) = \sum_{a=1}^{N} (-1)^a \Det (|_1 \overline{\psi^*}^a) \, ,
\end{equation}
where $(|_1 \overline{\psi^*}^a)$ denotes the $(N-1) \times (N-1)$ minor matrix obtained from the
$N \times N$ matrix $(\psi)$ by deleting the first column and the $a$-th row. The derivatives can
then be carried out:
\begin{widetext}
\begin{eqnarray}\label{momentumsquared}
\left<(p_1)_i(p_1)_j\right>  = \frac{1}{N!} \sum_{a,b=1}^N (-1)^{a+b}\sum_{\n,\m}
(c_\m)_a^*(c_\n)_b \left[(q_b)_i(q_b)_j+\hbar^2 n_l k_{il} n_k k_{jk} - \hbar n_k k_{ik} (q_b)_j
-\hbar n_l k_{jl} (q_b)_i\right]
\\ \times \frac 1V \int_V \exp\left\{i(x_1)_l\left[(m_i-n_i)k_{li}+\frac 1 \hbar
[(q_b)_l-(q_a)_l]\right]\right\} d^3x_1 \int_V \Det(|_1 \overline{\psi^*}^a) \Det(|_1
\overline{\psi}^b)d^3x_2 \ldots d^3x_N \, .\nonumber
\end{eqnarray}
\end{widetext}
Note that only the first integral in Eq. (\ref{momentumsquared}) contains $\vec x_1$. Using the
abbreviation
\begin{equation}
\kappa_l =(m_i-n_i)k_{li}+\frac 1 \hbar [(q_b)_l-(q_a)_l] \, ,
\end{equation}
the $d^3x_1$-integration in Eq. (\ref{momentumsquared}) can be expressed as
\begin{eqnarray}\label{dxeinsintegral}
\int_V \exp \left\{i(x_1)_l \kappa_l \right\} d^3x_1 = \left\{
\begin{array}{cc} V & (\kappa_l = 0) \, , \\ 0 & (\kappa_l \neq 0)  \, . \end{array}
\right.
\end{eqnarray}

Since the quasi-momenta $(q_a)_i$ are within the first Brillouin zone (\cite{Ashcroft}, p. 89),
$|(q_a)_l| \leq (\hbar/2)|n_jk_{lj}|$ for any $n_j \in \{\mathbb Z \backslash 0\}$. Thus,
$\kappa_l=0$ only if $n_i = m_i$ and $(q_b)_l = (q_a)_l$. We may assume that this holds only for
$a= b$. That allows to carry out the $\m$ and the $b$ summations. We now use
\begin{equation}
\sum_\n |(c_\n)_a|^2 n_i = 0 \, ,
\end{equation}
which follows from Eq. (\ref{cminusn}) and eliminates the terms linear in $n$ from the first line
of Eq. (\ref{momentumsquared}) \cite{noinversionsymmetry}. We define
\begin{equation}\label{xilk}
(\xi_a)_{lk} := \sum_\n |(c_\n)_a|^2 n_l n_k
\end{equation}
and
\begin{equation}
\overline{q_iq_j}:=\frac 1N\sum_a (q_a)_i(q_a)_j \, ,
\end{equation}
the average of the quasi-momentum product $q_iq_j$ over all electrons. We also define the average
\begin{equation}\label{overlinexilk}
\overline{\xi_{ij}}:=\frac 1N \sum_a (\xi_a)_{ij} \, .
\end{equation}
Together with Eq. (\ref{cnnorm}), this yields
\begin{eqnarray}
\left<(p_1)_i(p_1)_j\right> = \frac{N}{N!} \left[\overline{q_iq_j}+\hbar^2
\overline{\xi_{lk}}k_{il} k_{jk} \right] \nonumber \\ \times \int_V d^3x_2 \ldots d^3x_N \Det (|_1
\overline{\psi^*}^a) \Det (|_1 \overline{\psi}^b)
\\ = \left[\overline{q_iq_j}+\hbar^2 \overline{\xi_{lk}}k_{il}
k_{jk} \right] \left<\Psi^{N-1}|\Psi^{N-1}\right> \, . \nonumber
\end{eqnarray}
To prove that $\left<\Psi^{N-1}|\Psi^{N-1}\right>=1$ we expand the remaining determinants of $(|_1
\overline{\psi^*}^a)$ in terms of the column which is now the first one. The $d^3x_2$ integration
can then be carried out. The procedure is repeated, until the $d^3x_N$ integration is done. Each
step reduces the dimension of the Slater determinant by one and produces a factor equal to the
number of electrons still involved. Taking all these factors together cancels the normalization
factor $1/(N-1)!$. Thus, we obtain the desired result
\begin{equation}\label{pjquadrat}
\left<(p_1)_i(p_1)_j\right> = \overline{q_iq_j}+\hbar^2 \overline{\xi_{lk}}k_{il} k_{jk} \, .
\end{equation}
Since the right hand side of this result contains no reference to the electron's number, it holds
for all $a=1, \ldots, N$ electrons.

\subsubsection{Estimating $\overline{\xi_{lk}}$}

The properties of the wave function enter the momentum expectation value via $\overline{\xi_{lk}}$
and $\overline{q_iq_j}$.

$\overline{\xi_{lk}}$ is an ensemble average over all electrons:
\begin{equation}\label{xilkaverage}
\overline{\xi_{lk}}=\sum_\n \left(\frac 1N \sum_a |(c_\n)_a|^2 n_ln_k \right)=: \sum_\n
\overline{|c_\n|^2} n_ln_k \, ,
\end{equation}
obtained by substituting Eq. (\ref{xilk}) into Eq. (\ref{overlinexilk}). A detailed evaluation of
$\overline{\xi_{lk}}$ would start from material specific Fourier coefficients $c_\n$ obtained
experimentally or theoretically (see, e.g, \cite{JonesMarch}). Since detailed wave-function
calculations for realistic materials are notoriously difficult and would have to be performed for
each individual material, it is interesting to use a relatively simple model for the Fourier
coefficients. Such a model might already be quite accurate, since only the average of the absolute
square $\overline{|c_\n|^2}$ is required, rather than the detailed $(c_\n)_a$ for the individual
electrons. A (possibly complicated) dependency of the $(c_\n)_a$ on $\n$ can be hoped to smooth
out in the averaging. The model must, however, be in accordance with the requirement that the wave
function has the rotational symmetry of the lattice, i.e, $c_{\hat \rho \n} \equiv c_\n$ if $\hat
\rho $ is any operator of the rotation group of the crystal (\cite{JonesMarch}, pp. 469).

For such a simple model, we assume that $\overline{|c_\n|} \equiv
 \overline{|c_{|\n |}|} $, i.e., the average of the absolute squares of the
average of the Fourier coefficients depends only on $|\n |$. It follows that
\begin{equation}\label{xilkaveragemodell}
\overline{\xi_{lk}}=\gamma_{\rm mat} \delta_{lk}
\end{equation}
with some material dependent constant $\gamma_{\rm mat}$. For determining $\gamma_{\rm mat}$, we
note that the average kinetic energy of an electron $\left< T \right> = \frac{1}{2m} \left< p_i
p_j \right> \delta_{ij}$. From Eqs. (\ref{pjquadrat}), (\ref{xilkaverage}), and
(\ref{xilkaveragemodell}),
\begin{equation}\label{ekin}
\left< T \right> = \gamma_{\rm mat}\frac{ \hbar^2}{2 m} \delta_{lk} k_{il} k_{jk} \delta_{ij} \, .
\end{equation}
If we neglect for the moment the energy of the chemical bonding, this should correspond to the
average of the kinetic energies of the atoms' valence electrons, which can be estimated using the
Bohr model. The kinetic energy of an electron in the Bohr model $T_{\rm Bohr}= E_R Z/n^2$. Here,
$E_R \simeq 13.6$\,eV is the Rydberg energy, $Z$ the charge number of the atom core, and $n$ the
principal quantum number. Thus, the kinetic energy from the Bohr model, averaged over the
electrons within the atoms of the molecule,
\begin{equation}
\overline{T_{\rm Bohr}} = \frac{1}{N_{e,m}} \sum_{k=1}^{N_{e,m}} v_k \frac{E_R Z_k}{n_k^2}
\end{equation}
where $N_{e,m}$ gives the number of valence electrons per molecule. The index $k$ enumerates the
atoms of the molecule and $v_k, n_k$, and $Z_k$ are the valence, principal quantum number, and
charge number, respectively, of the atom $k$. Note that $Z_k = v_k$, since in an atom with $v$
valence electrons, the charge number of the atom cores and the inner shell electrons is $Z= v$.
For example, in quartz, SiO$_2$, there is one Si atom with $v=4$, $n=3$ and two O atoms with $v=2$
and $n=2$. Thus, for example,
\begin{equation}
\overline{(T_{\rm Bohr})}_{\rm quartz} = \frac{E_R}{8} \left( 1 \frac{4^2}{3^2}+ 2 \frac{2^2}{2^2}
\right) = \frac{17}{36} E_R \, .
\end{equation}
Comparing this to Eq. (\ref{ekin}), we obtain the material specific values $\gamma_{\rm mat}$
given in Tab. \ref{matb11}.

The model can be refined by taking into account the energy of the chemical bonding, which leads to
an increase of the actual kinetic energy of the electrons. The so called enthalpy of formation
$\Delta_f H^0$ gives the enthalpy for the formation of the crystal from the elements in their
usual state at standard conditions (room temperature and pressure), e.g., solid or diatomic
(O$_2$, for example). The Bohr model, however, predicts the energy of the unbound atoms. That
means, the change $T-T_{\rm Bohr}$ between the sum of the kinetic energy of the valence electrons
of the free atoms $T_{\rm Bohr}$ and the sum of their kinetic energy in the molecule $T$ is the
difference of $\Delta_f H^0$ and the applicable enthalpies of sublimation $\Delta_{\rm subl} H^0$
or dissociation $\Delta_{\rm diss} H^0$ of the elements. For sapphire, Al$_2$O$_3$, for example,
\[
\begin{array}{ccc}
\Delta_f H^0(\mbox{Al}_2\mbox{O}_3) & = & 16.8\,\mbox{eV} \\
-\Delta_{\rm subl} H^0 (2\mbox{Al}) &
= & -2\times 3.0\,\mbox{eV}
\\ -\Delta_{\rm diss} H^0(3\mbox{O}) & = & -3\times 2.5\,\mbox{eV}
\\  \hline T-T_{\rm Bohr} & = & 3.3\, \mbox{eV} \, ,
\end{array}
\]
or about 5\% of $T_{\rm Bohr}=68$\,eV. This would lead to a 5\% increase of the factor
$\gamma_{\rm mat}$. This indicates that the energy of the chemical bonding is, for our purposes,
negligible.

The ultimate refinement of the model would be the insertion of material specific Fourier
coefficients $c_\n$ into Eq. (\ref{xilkaverage}). The precision of the model would then approach
the limitations of the Bloch ansatz for the wave function itself, which is based on a mean field
model for the electron-electron interactions. Such a detailed model is, however, beyond the scope
of the present work.

\subsubsection{Result}

The expectation value of the Lorentz-violating correction to the single particle hamiltonian
$\sum_a \delta h_a$ can be obtained from Eq. (\ref{pjquadrat}) by multiplying with the number of
electrons. Inserting $\overline{\xi_{lk}}$ as obtained in the previous section into Eq.
(\ref{pjquadrat}), we obtain
\begin{equation}
\left< \delta h \right>  = \frac N m E'_{ij} \left( \overline{q_iq_j} + \gamma_{\rm mat} \hbar^2
\delta_{lk} k_{il} k_{jk} \right) \, .
\end{equation}

The quasi-momenta $|\vec q_a|$ are restricted to the first Brillouin zone, $|\vec q| < \frac \hbar
2 |\vec k_{j}|$. Most electrons, however, will have a quasi-momentum lower than this maximum
value, so that the average $\overline{q_i q_j}$ is a relatively small contribution to $\left<
\delta h \right>$. If we neglect it, we obtain the final result
\begin{eqnarray}\label{deltahresult}
\left< \delta h \right>  =  N \gamma_{\rm mat} \frac{\hbar^2}{m} E'_{ij} \delta_{lk} k_{il} k_{jk}
\, .
\end{eqnarray}
$\delta_{lk} k_{il} k_{jk}$ is symmetric in the indices $i$ and $j$. This result will be used in
the next section to compute the geometry change of the crystal caused by Lorentz violation in the
electrons' equation of motion.

\section{Change of crystal geometry}\label{lengthchange}

The direct lattice vectors $\vec l_a$ contained in the matrix $l_{ia}$ determine the structure of
the lattice without Lorentz violation. Lorentz violation will cause a change $\tilde l_{ia}$ of
the crystal geometry, i.e., the lattice vectors are now given by $l_{ia}+ \tilde l_{ia}$. To
calculate it, we adjust $\tilde l_{ia}$ such as to minimize the total energy of the lattice
\begin{equation}
U= U_0 (l_{ia}+\tilde l_{ia}) +\left<\delta h(l_{ia}+\tilde l_{ia}) \right> \, .
\end{equation}
The first term is the conventional total energy of the lattice without Lorentz violation. It can
be expressed as
\begin{equation}
U_0 (l_{ia}+\tilde l_{ia}) = U_0(l_{ia}) + U_{\rm elast}(\tilde l_{ia})+U_c
\end{equation}
where $U_c$ is a constant and $U_{\rm elast}$ is the elastic energy connected to a distortion of
the lattice. If $\tilde l_{ia} =0$, the elastic energy $U_{\rm elast}=0$. The total energy is thus
given by
\begin{equation}
U = U_0(l_{ia}) + U_{\rm elast}(\tilde l_{ia})+ \left<\delta h(l_{ia}+ \tilde l_{ia}) \right> \,.
\end{equation}
The correction $\tilde l_{ia}$ is found by setting to zero the derivative:
\begin{equation}
\frac{\partial U}{\partial \tilde l_{jb}} =\frac{\partial U_{\rm elast} }{\partial \tilde l_{jb}}
+ \frac{\partial \left<\delta h(l_{ia}+ \tilde l_{ia}) \right>}{\partial \tilde l_{jb}} = 0 \,.
\end{equation}
To explicitly calculate $\tilde l_{jb}$, we have to express the contributions to $U$ in terms of
$\tilde l_{jb}$. We will do so for $\left< \delta h \right>$ in Sec. \ref{lorentzviolationpart}
and in Sec. \ref{elenergy} for $U_{\rm elast}$. In Sec. \ref{minenerg}, the total energy per unit
cell thus obtained is minimized and the geometry change as expressed by a strain tensor $e_{ij}$
is calculated.


\subsection{Dependence of $\delta h$ on $\tilde l_{ia}$}\label{lorentzviolationpart}

The change of the hamiltonian's expectation value $\left< \delta h \right>$, Eq.
(\ref{deltahresult}) depends on the geometry via the reciprocal lattice vectors $k_{ij}$, for
which we have the relation (\cite{Ashcroft}, p. 87),
\begin{equation}\label{kdefinition}
l_{ij} k_{ik} =2\pi \delta_{jk} \, ,
\end{equation}
and therefore $l_{ij}k_{nj}=2\pi\delta_{in}$. If we substitute $l_{ij}+\tilde l_{ij}$ and $k_{jk}
+ \tilde k_{jk}$, with $\tilde l_{ij} \ll l_{ij}$ and $\tilde k_{jk} \ll k_{jk}$, we obtain
\begin{equation}
(l_{ij}+\tilde l_{ij})(k_{ik} + \tilde k_{ik})=2\pi \delta_{jk}
\end{equation}
or
\begin{equation}
l_{ij} k_{ik}+l_{ij} \tilde k_{ik}+ \tilde l_{ij}k_{ik}+\tilde l_{ij}\tilde k_{ik}=2\pi
\delta_{jk}\, .
\end{equation}
The first term on the l.h.s. cancels with the r.h.s due Eq. (\ref{kdefinition}); we neglect the
second order term on the l.h.s., and obtain
\begin{equation}
l_{ij} \tilde  k_{ik}+\tilde l_{ij} k_{ik}=0\, .
\end{equation}
Multiplying with $k_{nj}$ and using Eq. (\ref{kdefinition}) again,
\begin{equation}
\tilde k_{nk}=-\frac{1}{2\pi}k_{nj} \tilde l_{ij} k_{ik} \, .
\end{equation}
We can now substitute $k_{ij} + \tilde k_{ij}$ into Eq. (\ref{deltahresult}) to obtain $\delta
h(\tilde l_{ab})$:
\begin{eqnarray}
\delta h (\tilde l_{ab}) & = & N \gamma_{\rm mat} \frac{ \hbar^2}{m} \delta_{lk} \left(k_{il}-
\frac{1}{2\pi} k_{ia} \tilde l_{ba} k_{bl}\right)  \nonumber \\ & & \times
\left(k_{jk}-\frac{1}{2\pi} k_{ja} \tilde l_{ba} k_{bk}\right) E'_{ij}
\\ & = & \mbox{const} - \gamma_{\rm mat} \frac{N \hbar^2}{2\pi m}
E'_{ij} \\ & & \times  (k_{ia}\tilde l_{ba} k_{bk}k_{jk}+k_{ja}\tilde l_{ba}k_{bk}k_{ik}) \, .
\nonumber
\end{eqnarray}
The const$ = N \gamma_{\rm mat} \frac{\hbar^2}{m} E'_{ij} \delta_{lk} k_{il}k_{jk}$ does not
depend on $\tilde l_{ab}$. A term of order $E'_{ij}(\tilde l_{ab})^2$ has been neglected.


\subsection{Elastic energy}\label{elenergy}

The elastic energy is given as \cite{Woan}
\begin{equation}
U_{\rm elast}=\frac 12 \lambda_{ijkl}e_{ij} e_{kl} V \, ,
\end{equation}
where $\lambda_{ijkl}$ is the elastic modulus, $V$ the volume considered, and
\begin{equation}
e_{ij}=\frac 12 \left(\frac{\partial u_i}{\partial x_j}+\frac{\partial u_j}{\partial x_i} \right)
\end{equation}
is the strain tensor, where $u_i$ is the displacement of a volume element at some location $x_i$.
For $i=j$, $u_{ij}$ represents the relative change of length in $x_i$-direction, and for $i \neq
j$, it represents the change of the right angle between lines originally pointing in $x_i$ and
$x_j$ direction.

To express the elastic energy in terms of $\tilde l_{ij}$, we note that the location $x_b$ of a
point of the direct lattice can be expressed as a linear combination of the primitive lattice
vectors
\begin{equation}
x_b=\eta_a l_{ba} \, ,
\end{equation}
with some coefficients $\eta_a \in \mathbb{N}$. Multiplying this equation by $k_{bc}$ and using
Eq. (\ref{kdefinition}), we obtain
\begin{equation}
\eta_c= \frac{1}{2\pi} k_{bc} x_b \, .
\end{equation}
If $l_{ba}$ is shifted to $l_{ba}+ \tilde l_{ba}$, the lattice point originally at $x_b$ will be
shifted to $x_b + u_b$, where
\begin{equation}
u_b= \eta_a \tilde l_{ba} = \frac{1}{2\pi} k_{da} x_d \tilde l_{ba} \, .
\end{equation}
Therefore, we have
\begin{equation}
\frac{\partial u_d}{\partial x_c}= \frac{1}{2\pi}k_{ca}\tilde l_{da}
\end{equation}
or
\begin{equation}\label{straintensor}
e_{dc}=\frac{1}{4\pi}\left(k_{da} \tilde l_{ca} + k_{ca} \tilde l_{da} \right)\, .
\end{equation}
This can now be used to express the elastic energy in terms of $\tilde l_{ab}$:
\begin{equation}
U_{\rm elast}=\frac 12 \lambda_{ijkl} \frac{1}{4\pi}(k_{ia}\tilde l_{ja} + k_{ja} \tilde l_{ia})
\frac{1}{4\pi}(k_{lb} \tilde l_{kb} + k_{kb} \tilde l_{lb}) V \, .
\end{equation}
Some manipulation of indices using $\lambda_{abcd}=\lambda_{bacd}=\lambda_{cdab}$ leads to the
more simple form
\begin{equation}
U_{\rm elast}=\frac{V}{8\pi^2} \lambda_{ijkl} k_{ia} \tilde l_{ja} k_{lb} \tilde l_{kb} \, .
\end{equation}


\subsection{Minimizing the total energy per unit cell}\label{minenerg}

Summing up the contributions, we find for the energy change per unit cell (leaving out the
constant terms)
\begin{eqnarray}
U & = & |\mbox{det}(l_{ij})|\frac{1}{8\pi^2} \lambda_{ijkl} k_{ia} \tilde l_{ja} k_{lb} \tilde
l_{kb} - \gamma_{\rm mat} \frac{N \hbar^2}{2\pi m} \\ & & \times E'_{ij} (k_{ia}\tilde l_{ba}
k_{bk}k_{jk}+k_{ja}\tilde l_{ba}k_{bk}k_{ik}) \nonumber
\end{eqnarray}
were $V=|$det$(l_{ij})|$, the volume of a unit cell, and $N=N_{e,u}$ the number of valence
electrons per unit cell have been inserted (not to be confused with $N_{e,m}$ the corresponding
number per molecule). The inner-shell electrons are assumed not to influence the crystal geometry.
A minimum is found, when
\begin{equation}
\frac{\partial U}{\partial \tilde l_{mn}}=0 \, .
\end{equation}
After some manipulation, the derivative can be expressed as
\begin{eqnarray}
\frac{\partial U}{\partial \tilde l_{mn}} & = & \frac{|\mbox{det}(l_{ij})|}{4\pi^2} \lambda_{imkl}
k_{in} k_{lb} \tilde l_{kb} -
 \gamma_{\rm mat}\frac{N_{e,u} \hbar^2}{2 \pi m}
 \\ & & \times E'_{ij}
(k_{in} k_{mk} k_{jk} + k_{jn} k_{mk} k_{ik}) = 0 \, .\nonumber
\end{eqnarray}
We denote $E'_{(ij)}= \frac 12 (E'_{ij}+E'_{ji})$ the symmetric part of the tensor $E'_{ij}$. The
last equation can be simplified a bit by multiplying with $l_{pn}$:
\begin{equation}\label{tildelroh}
\lambda_{pmkl} k_{lb} \tilde l_{kb}- \gamma_{\rm mat} \frac{ 4 \pi
\hbar^2N_{e,u}}{|\mbox{det}(l_{ij})| m} E'_{(pj)} k_{mk} k_{jk} = 0 \, .
\end{equation}
For solving this for $\tilde l$, we need the inverse $\mu_{abkl}$ (called the compliance tensor)
of $\lambda_{abcd}$, defined by
\begin{equation}
\mu_{abkl} \lambda_{abcd} = \delta_{kc} \delta_{dl} \, .
\end{equation}
Multiplying Eq. (\ref{tildelroh}) with $\mu_{depm}$ gives
\begin{equation}
k_{eb} \tilde l_{db} =\gamma_{\rm mat} \frac{4\pi N_{e,u}}{|\mbox{det}(l_{ij})|}
 \frac{\hbar^2}{m} E'_{(pj)}\mu_{demp} k_{mk}k_{jk}
\end{equation}
and a further multiplication by $l_{es}$ yields
\begin{equation}
\tilde l_{ds}=\gamma_{\rm mat} \frac{2\hbar^2 N_{e,u}}{m| \mbox{det}(l_{ij})|} E'_{(pj)}
\mu_{demp} l_{es} k_{mk} k_{jk} \, .
\end{equation}
The strain tensor can be calculated from this result using Eq. (\ref{straintensor}) as
\begin{equation}\label{edctensor}
e_{dc} = \tilde \mathcal B_{dcpj} E'_{(pj)}
\end{equation}
with
\begin{equation}
\tilde \mathcal B_{dcpj}=  \gamma_{\rm mat}\frac{2 N_{e,u}  \hbar^2}{m |\mbox{det}(l_{ij})|}
\mu_{dcmp} k_{mk} k_{jk} \, .
\end{equation}
This has been simplified by using Eq. (\ref{kdefinition}). Since for Eq. (\ref{edctensor}), this
is multiplied with the symmetric $E'_{(jp)}$, only the part $\mathcal B_{dcpj} = \frac 12 (\tilde
\mathcal B_{dcpj} + \tilde\mathcal B_{dcjp})$ that is symmetric in $j$ and $p$
\begin{equation}\label{bdefinition}
\mathcal B_{dcjp} =  \gamma_{\rm mat}\frac{N_{e,u}  \hbar^2}{m |\mbox{det}(l_{ij})|}
\left(\mu_{dcmp} k_{mk} k_{jk}+\mu_{dcmj} k_{mk} k_{pk}\right)
\end{equation}
is used. The resulting strain tensor
\begin{equation}
e_{dc}=\mathcal B_{dcpj} E'_{pj}
\end{equation}
is given by the $3\times 3$ Lorentz violation tensor $E'_{pj}$ and a tensor $\mathcal B_{dcpj}$,
which gives the sensitivity of the material geometry change to Lorentz violation in the electron
sector of the SME. This is the desired result of this section.

\subsubsection{General properties of the sensitivity tensor $\mathcal B_{dcpj}$ and conventions}
The sensitivity tensor has the symmetries
\begin{equation}
\mathcal B_{dcjp} = \mathcal B_{cdjp} = \mathcal B_{dcpj} \, .
\end{equation}
In general, however, $\mathcal B_{dcjp} \neq \mathcal B_{jpdc}$. It has, therefore, at most 36
independent elements, the number of which is reduced for a symmetric crystal. For a compact
presentation of the material specific results in the following sections, we will arrange these
into a $6\times 6$ matrix, that allows to express Eq. (\ref{edctensor}) as a 6 dimensional matrix
equation
\begin{equation}\label{lchange6times6}
e_\Gamma = \mathcal B_{\Gamma \Xi} E'_\Xi \,.
\end{equation}
We therefore arrange the six independent elements of $e_{dc}$ and $E'_{(bj)}$ as the vectors
\begin{eqnarray}
e_\Gamma & = & (e_{xx}, e_{yy}, e_{zz}, e_{yz}, e_{zx}, e_{xy}) \, , \\ E'_\Gamma & = & (E'_{xx},
E'_{yy}, E'_{zz}, E'_{yz}, E'_{zx}, E'_{xy})
\end{eqnarray}
[the capital greek indices run from $1 \ldots 6$] and define the sensitivity matrix
\begin{equation}\label{sensitivitymatrix}
\mathcal B=\left( \begin{array}{cccccc} \mathcal B_{1111} & \mathcal B_{1122} & \mathcal B_{1133}
& 2\mathcal B_{1123} & 2\mathcal B_{1131} & 2\mathcal B_{1112}
\\ \mathcal B_{2211} & \mathcal B_{2222} & \mathcal B_{2233} & 2\mathcal B_{2223} & 2\mathcal B_{2231} & 2\mathcal B_{2212} \\
\mathcal B_{3311} & \mathcal B_{3322} & \mathcal B_{3333} & 2\mathcal B_{3323} & 2\mathcal
B_{3331} & 2\mathcal B_{3312} \\ \mathcal B_{2311} & \mathcal B_{2322} & \mathcal B_{2333} &
2\mathcal B_{2323} & 2\mathcal B_{2331} & 2\mathcal B_{2312} \\ \mathcal B_{3111} & \mathcal
B_{3122} & \mathcal B_{3133} & 2\mathcal B_{3123} & 2\mathcal B_{3131} & 2\mathcal B_{3112} \\
\mathcal B_{1211} & \mathcal B_{1222} & \mathcal B_{1233} & 2\mathcal B_{1223} & 2\mathcal
B_{1231} & 2\mathcal B_{1212} \\
\end{array} \right) \, .
\end{equation}
The factors of 2 account for the double-counting of the non-diagonal elements of $E'_{(pj)}$ in
the tensor equation Eq. (\ref{edctensor}).

\subsubsection{Sensitivity matrix for isotropic materials}
Let us first consider isotropic materials that have no preferred crystal orientation, i.e.
crystals of cubic structure and non--crystalline (fused) materials which consist of a large number
of small crystals oriented statistically. Cubic materials have one single lattice constant $a$;
the matrix of the primitive direct lattice vectors is given by $l_{ij} = a \delta_{ij}$. According
to Eq. (\ref{kdefinition}), the matrix of the reciprocal lattice vectors is given by $k_{ij}=(2
\pi /a)\delta_{ij}$. In the appendix, it is described how to obtain the compliance constants
$\mu_{abcd}$ from the elasticity constants that are tabulated for various materials in the
literature, e.g. \cite{CRC}. Inserting into Eqs. (\ref{bdefinition},\ref{sensitivitymatrix}), we
obtain the sensitivity matrix $\mathcal B$. It is of the structure
\begin{equation}\label{matrixcubic}
\mathcal B= \left( \begin{array}{cccccc} \mathcal B_{11} & \mathcal B_{12} & \mathcal B_{12} & 0 &
0 & 0
\\ \mathcal B_{12} & \mathcal B_{11} & \mathcal B_{12} &0 & 0& 0\\ \mathcal B_{12} & \mathcal
B_{12} & \mathcal B_{12} & 0& 0& 0\\ 0 & 0 & 0 & \mathcal B_{44} & 0& 0\\0 & 0 & 0 & 0 & \mathcal
B_{44} & 0
\\ 0 & 0 & 0 & 0 & 0 & \mathcal B_{44}
\end{array} \right) \, .
\end{equation}
For cubic crystals, the non-zero values
\begin{eqnarray*}
\mathcal B_{11} & = & \xi /(a^2) C_{11} \, , \\ \mathcal B_{12} & = & \xi/(a^2) C_{12} \, , \\
\mathcal B_{44} & = & \xi/(2a^2) C_{44} \, ,
\end{eqnarray*}
where
\begin{equation}
\xi =  \gamma_{\rm mat}\frac{8 \pi^2 N_{e,u}  \hbar^2}{m |\mbox{det}(l_{ij})|}
\end{equation}
and $C_{\Gamma \Xi}$ are the elements of the $6 \times 6$ compliance matrix, Eq.
(\ref{compliancematrix}). From symmetry arguments, this is also the structure of the $\mathcal B$
matrix for non crystalline materials without a preferred orientation. The elements of this matrix
for some cubic and/or fused materials are given in table \ref{matb11}. The values for fused quartz
and sapphire have been calculated from the values of the crystalline materials (to be calculated
below) as averages over crystal orientations.

\subsubsection{Sensitivity matrix for trigonal crystals}

Quartz and Sapphire are of trigonal structure and are frequently used in cavity experiments.
Therefore, we also consider the trigonal case here. The matrix of the primitive direct lattice
vectors can be chosen as
\begin{equation}
l_{ij} = \left( \begin{array}{ccc} a/2 & a/2 & 0
\\ \sqrt{3}a/2 & -\sqrt{3}a/2 & 0 \\ 0 & 0 & c \end{array} \right) \,
\end{equation}
where $a$ and $c$ are the two lattice constants. We calculate the inverse $k_{ij}$; the product
\begin{equation}
k_{ik} k_{jk} = 4 \pi^2 \left( \begin{array}{ccc} 2/a^2 & 0 & 0 \\ 0 & 2/(3a^2) & 0 \\ 0 & 0 &
1/c^2
\end{array} \right)
\end{equation}
turns out to be a diagonal matrix. Trigonal crystals have six independent compliance constants and
two lattice constants, which makes 8 independent components for the $\mathcal B$-matrix. It has
the structure
\begin{equation}\label{matrixtrigonal} \mathcal B = \left(
\begin{array}{cccccc}
  \mathcal B_{11} & \mathcal B_{12} & \mathcal B_{13} & \mathcal B_{14} & 0 & 0 \\
  3\mathcal B_{12} & \frac 13 \mathcal B_{11} & \mathcal B_{13} & -\mathcal B_{14} & 0 & 0\\
  \mathcal B_{31} & \frac 13 \mathcal B_{31} & \mathcal B_{33} & 0 & 0 & 0 \\
  \mathcal B_{41} & -\frac 13 \mathcal B_{41} & 0 & \mathcal B_{44} & 0 & 0\\
  0 & 0 & 0 & 0 & \mathcal B_{55} & \frac 23 \mathcal B_{41} \\
  0 & 0 & 0 & 0 & \mathcal B_{55} & \mathcal B_{66}
\end{array} \right)
\end{equation}
with
\begin{eqnarray*}
\mathcal B_{55} & = & \mathcal B_{44}+\frac 13 \mathcal B_{41} \, , \\ \mathcal B_{66} & = & \frac
13\mathcal B_{11}-\mathcal B_{12} \, .
\end{eqnarray*}
The matrix elements are explicitly
\begin{eqnarray*}
\mathcal B_{11} & = & 2\xi/(a^2) C_{11} \, , \\ \mathcal B_{12} & = & 2\xi/(3a^2) C_{12} \, , \\
\mathcal B_{13} & = & \xi/(c^2) C_{13} \, , \\ \mathcal B_{14} & = &\xi [1/(3a^2)+1/(2c^2)]C_{14}
\, , \\ \mathcal B_{31} & = & 2\xi/(a^2) C_{13} \, ,
\\ \mathcal B_{33} & = & \xi/(c^2) C_{33} \, ,
\\ \mathcal B_{41} & = & \xi/(a^2) C_{14} \, , \\ \mathcal B_{44} & = & \xi[1/(6a^2)+1/(4c^2)] C_{44} \, .
\end{eqnarray*}
$C_{\Gamma \Xi}$ are the elements of the compliance matrix given in Eq. (\ref{compliancematrix}).
Numerical values of $\mathcal B_{\Gamma \Xi}$ for quartz and sapphire are given in Tab.
\ref{matb11}. The matrix $\mathcal B$ is not symmetrical; the elements of the first column are
generally the highest in this matrix, i.e., the geometry change of trigonal materials is most
sensitive to the $xx$ element of the Lorentz violation parameters $E'_{(ij)}$. This is because the
direct lattice vector components in $x$-direction are a factor $\sqrt{3}$ smaller than the $y$
components. Hence, the wave function of the electrons oscillate faster in $x$ direction, i.e. the
$\left<p_x\right>$ momentum component is larger. Since the influence of Lorentz violation is given
by the $\left<p_i p_j\right>$ matrix element, this means a higher influence of the $x$-component
of the Lorentz violation coefficients $E'_{(ij)}$.

High elastic constants decrease the values of $\mathcal B$ so that crystals of high stiffness
(such as sapphire) should show lower values of $\mathcal B$. However, in some cases (particularly
diamond), this is outweighed by small dimensions of the unit cell (that imply high momentum
expectation values due to short period of the electron wave functions) and a large number of
electrons per unit cell.

\begin{table}[t]
\caption{Elements of the sensitivity matrix for fused and/or cubic materials. fq denotes fused
quartz, fs fused sapphire, C denotes diamond. Materials for which three elements of $\mathcal B$
are given are isotropic; the coefficients should be inserted into Eq. (\ref{matrixcubic}). The
other materials are trigonal; the coefficients for these are to be inserted into Eq.
(\ref{matrixtrigonal}). \label{matb11}}

\begin{tabular}{lccccccccc}

\hline

Mat. & $\gamma_{\rm mat}$ & $\mathcal B_{11}$ & $\mathcal B_{12}$ & $\mathcal B_{13}$ & $\mathcal
B_{14}$ & $\mathcal B_{31}$ & $ \mathcal B_{33}$ & $\mathcal B_{41}$ & $\mathcal B_{44}$ \\ \hline

fq & 0.38 & 0.77 & -0.09   &        - & -        & - & - & - & 0.57
\\

fs & 0.29 & 0.06 & -0.01 & -        & -        & - & - & - & 0.05
\\

Si & 0.50 & 2.51     & -0.70     & -        & -        & - & - & - & 2.05 \\

C & 1.16 & 5.77    & -0.59     & -        & -        & - & - & - & 5.35 \\ \hline

Al$_2$O$_3$ & 0.29 & 0.14   & -0.01    & -0.00    & 0.02    & -0.02 & 0.01 & 0.01 & 0.04
\\

SiO$_2$  & 0.38 & 1.41   & -0.07     & -0.06     & 0.35  &-0.14 & 0.44 & 0.25 & 0.41
\\ \hline
\end{tabular}
\end{table}


\section{Experiments}\label{experiments}

Here, we discuss the application of our results to extract limits on Lorentz violation in the
electrons' equation of motion from experiments.

\subsection{Lorentz violation signal in cavity experiments}

As discussed in the introduction, Lorentz violation may affect the resonance frequency of a cavity
$\nu$ through a variation of $c$ and $L$:
\begin{equation}
\frac{\delta \nu}{\nu_0} = \frac{\delta c}{c_0} - \frac{\delta L}{L_0} \, .
\end{equation}
Here, $\nu_0, c_0$, and $L_0$ are quantities in absence of Lorentz violation. For cavity
experiments, we have to consider both Lorentz violation in electrodynamics as well as in the
electrons' equation of motion. The influence of Lorentz violation in electrodynamics leads to a
frequency change $\delta_{\rm EM} \nu$ that is mainly caused by a variation of the velocity of
light, $\delta_{\rm EM} c$ \cite{KosteleckyMewesPRL,KosteleckyMewesPRD}, and a small
material-dependent contribution due a length change $\delta_{e-} L$, that is usually negligible
\cite{TTSME}. Lorentz violation in the electrons' equation of motion affects solely $L$. This
leads to a frequency change $\delta_{e-} \nu$, so that
\begin{equation}
\delta \nu = \delta_{\rm EM} \nu + \delta_{e-} \nu \, .
\end{equation}
For example, if the cavity axis is parallel to the $z$-axis of the crystal,
\begin{equation} -\frac{\delta_{e-} \nu}{\nu_0} =
\frac{\delta L_z}{L_{z,0}} \equiv e_3 \, .
\end{equation}
From Eq. (\ref{lchange6times6}), $e_3 = \mathcal B_{3\Xi} E'_\Xi$.

\subsection{Limits from previous experiments}\label{limitsfrompreviousexperiments}

Cavity experiments have been performed repeatedly as tests of Lorentz violation in electrodynamics
\cite{BrilletHall,HilsHall,Braxmaier,Wolf,Lipa,MuellerASTROD,MuellerMM}. It is interesting to
estimate the level of the limits on $c_{\mu \nu}$ resulting from these experiments. Since none of
them were done with a setup optimized for obtaining separate bonds on Lorentz violation in
electrodynamics and in the electrons' equation of motion (see below), we have to use an assumption
that simplifies the analysis, so we can work with the published data only. To obtain a sharp upper
limit, we have to compare two experiments of high precision that used different cavity materials.

M\"{u}ller {\em et al.} \cite{MuellerMM} performed a Michelson-Morley experiment using two
sapphire (s) cavities subject to Earth's rotation. The cavity axes were parallel to the crystals
$c$-axis. Since for such cavities, the elements of $\mathcal B_{3\Xi}^s$ giving the sensitivity to
Lorentz violation in the electronic sector are relatively low, we neglect the effect on Lorentz
violation in the electrons' equation of motion for this experiment, i.e., we view the experiment
as a pure test of Lorentz violation in electrodynamics. Due to Earth's rotation with the angular
frequency $\omega_\oplus \simeq 2 \pi/$23\,h56\,min and Earth's orbit with $\Omega_\oplus = 2
\pi/$1\,year, Lorentz violation in electrodynamics leads to a time-dependency of the frequency
difference between the cavities that has Fourier components at 6 frequencies that are linear
combinations of $\omega_\oplus$ and $\Omega_\oplus$. The experiment gives individual bounds on the
amplitudes of these Fourier components. At $2 \omega_\oplus$ and $2 (\omega_\oplus \pm
\Omega_\oplus)$, the bounds are below $4 \times 10^{-15}$.

Brillet and Hall \cite{BrilletHall} used a single fused quartz (fq) cavity (actually, ``ultra-low
expansion" glass ceramics, ULE) on a turntable rotating at $\omega_t$. Frequency measurement was
accomplished by comparison to a stationary CH$_4$ standard. Since in a single-cavity setup, the
same Lorentz violation in electrodynamics leads to half the frequency variation compared to a
two-cavity setup, the experiment of M\"{u}ller {\em et al.} excludes signals from Lorentz
violation in electrodynamics larger than $2 \times 10^{-15}$ for this experiment. We may thus view
this experiment as a measurement of
\begin{equation}
\frac{\delta_{e-} \nu}{\nu_0} = \mathcal B_{\bar 3 \bar \Xi}^{fq} E'_{\bar \Xi} \, .
\end{equation}
The indices $\bar \Xi$ denote components in the cavity frame of reference. In accordance with
\cite{Kostelecky99,KosteleckyMewesPRL,KosteleckyMewesPRD}, we define the $x$ axis of the
laboratory frame as the north-south axis, the $y$ axis as the east-west axis, and the $z$ axis as
pointing upwards. The turntable rotated in the $xy$-plane \cite{BrilletHall}. We define the cavity
frame $\bar z$ axis parallel to the $z$ axis, the $\bar x$ axis parallel to the cavity axis. A
calculation of the hypothetical signal starts from transforming $c_{\mu \nu}$ as given in the
sun-centered standard frame into the laboratory frame, as described in
\cite{Kostelecky99,KosteleckyMewesPRL,KosteleckyMewesPRD}. A further rotation around the
laboratory $z$ axis gives the quantities $c_{\bar \mu \bar \nu}$ in the cavity frame, from which
$E'_{\bar i \bar j}$ follows from Eq. (\ref{hamiltonianparameters}). Due to the rotations, the
hypothetical signal becomes time-dependent and is given by (assuming that at $\To=0$, the cavity
axis coincides with the $X$ axis, and neglecting terms proportional to Earth's orbital velocity
$\beta_\oplus \sim 10^{-4}$)
\begin{eqnarray}
\frac{\delta \nu}{\nu} & = &  C(2,0,0) \cos 2 \omega_t T_\oplus  + S(2,2,0) \sin (2 \omega_t +
2\omega_\oplus ) T_\oplus  \nonumber \\ & & + C(2,2,0) \cos (2\omega_t + 2\omega_\oplus) T_\oplus
+ \mathcal{A}+ \mathcal O(\beta_\oplus^2) \, ,
\end{eqnarray}
where $\mathcal{A}$ denotes Fourier components at other frequencies, for which no experimental
results are published in \cite{BrilletHall}. The sine component at $2 \ot$ vanishes. The
coefficients are
\begin{eqnarray*}
C(2,0,0) & = & -\frac 14(\mathcal B_{11}^{fq}-\mathcal B_{12}^{fq}) \sin^2 \chi_B (c_{XX}+c_{YY} -
2c_{ZZ})
\\ & \simeq & -0.13\,(c_{XX}+c_{YY} - 2c_{ZZ}) \, , \\
S(2,2,0) & = & \cos^4\frac{\chi_B}{2} (\mathcal B_{11}^{fq}-\mathcal B_{12}^{fq}) c_{(XY)} \\ &
\simeq & 0.58\, c_{(XY)} \, ,
\\ C(2,2,0) & = & \frac 12 \cos^4\frac{\chi_B}{2} (\mathcal B_{11}^{fq}-\mathcal B_{12}^{fq}) (c_{YY}-c_{XX})\\
& \simeq & 0.29\, (c_{YY}-c_{XX}) \, .
\end{eqnarray*}
$\chi_B$ denotes the colatitude of Boulder, $\chi_B \simeq 50^\circ$. $\mathcal B_{11}^{fq}$ and
$\mathcal B_{12}^{fq}$ are given in Tab. \ref{matb11}. From the experiment, an amplitude at $2
\omega_t$ of $\sim 2 \times 10^{-13}$ is found, attributed to a slight tilt in the horizontal
alignment of the turntable. In principle, a hypothetical Lorentz violation signal at $2 \omega_t$
cannot be separated in the analysis from a signal caused by such tilt. However, if we consider an
exact cancellation between a strong Lorentz violation signal and a strong tilt signal improbable,
we obtain a rough order-of-magnitude bound on $|c_{XX}+c_{YY}-2c_{ZZ}|$ at a level of about
$10^{-12}$. The measured upper limit on the amplitude of the component of the hypothetical signal
at $2(\omega_t+ \omega_\oplus)$ (which can be distinguished from a tilt generated signal in the
Fourier analysis) is $4 \times 10^{-15}$. Adding in quadrature the maximum possible contribution
from Lorentz violation in electrodynamics according to the result of \cite{MuellerMM}, $2 \times
10^{-15}$, we obtain $\sqrt{S(2,2,0)^2+C(2,2,0)^2} \lesssim 4.5 \times 10^{-15}$. This gives the
limits
\begin{eqnarray}
|c_{(XY)}| & \lesssim & 8 \times 10^{-15} \, , \\ |c_{XX}-c_{YY}| & \lesssim &  1.6 \times
10^{-14} \, . \nonumber
\end{eqnarray}
Given the small magnitude of $\mathcal B_{3\Xi}^s$, these values would indeed lead to negligible
contributions to the experiment of M\"{u}ller {\em et al.}, so our above assumptions seem
reasonable.

\subsection{Optimized setups}

To obtain clean separate bounds on Lorentz violation in electrodynamics and the electrons'
equation of motion, without using the above assumptions, a dedicated experiment is desirable. One
could, for example, compare the resonance frequencies $\nu_{fq}$ and $\nu_s$ of a cavity made from
fused quartz and one from crystalline sapphire, with the cavity axis parallel to the crystal $z$
axis, for example. This seems to be a realistic scenario, since such cavities have been used in
experiments and proved to be of high stability
\cite{BrilletHall,HilsHall,Braxmaier,MuellerASTROD,MuellerMM,COREs,Young}. The signal for Lorentz
violation would be the frequency difference $\nu_s-\nu_{fq}$. If a parallel arrangement of
cavities would be chosen, the influence of Lorentz violation in electrodynamics could be
eliminated, and a clean bound on some components of $E'_{jk}$ could be extracted,
\begin{equation}
\frac{\nu_s-\nu_q}{\nu} = (\mathcal B_{\bar 3 \bar \Xi}^{fq}-\mathcal B_{\bar 3 \bar \Xi}^s)
E'_{\bar \Xi} \, .
\end{equation}
Here, $\nu \simeq \nu_s \simeq \nu_q$ is the average frequency. An accuracy level of below one
part in $10^{15}$ in frequency comparisons of cavities made from quartz \cite{Young} and sapphire
\cite{MuellerMM} has been demonstrated in the laboratory. Thus, placing bounds of a few parts in
$10^{15}$ on the components of $E'_{JK}$ that dominate the signal seems feasible. A contribution
of the time-components $c_{(0J)}$ to the signal arises if one takes into account the laboratory
velocity given by the velocity $\beta_\oplus \sim 10^{-4}$ of Earth's orbit and $0 \leq \beta_L
\lesssim 1.5 \times 10^{-6}$ due to Earths rotation (depending on the geographical latitude). The
Lorentz transformations between the sun--centered inertial reference frame and the laboratory
frame lead to additional Fourier components of the signal that are proportional to $c_{(0J)}$ and
either $\beta_\oplus$ or $\beta_L$. In a Fourier analysis of a sufficiently long timetrace, the
Fourier components can be resolved and individual limits on almost all components of $c_{\mu \nu}$
(only $c_{00}$ does not lead to time-dependent signals to first order in $\beta_\oplus$ or
$\beta_L$) can be expected, at or below about a part in $10^{15}$ for the dominating parameters
and to about a part in $10^{11}$ for the parameters that are suppressed by $\beta_\oplus$. Future
space experiments, for which a resolution of the frequency measurement of up to $10^{-18}$ is
projected \cite{OPTIS,SUMO}, might bound the dominating components of $E'_{jk}$ at the $10^{-18}$
level, and the suppressed components of $c_{\mu \nu}$ at the $10^{-14}$ level.

Instead of using different cavity materials, two cavities made from the same crystalline material,
but having different orientations of the cavities with respect to the crystal axes might be used.
Using, e.g., quartz, a comparison between a cavity fabricated such that the cavity axis is
parallel to the crystals $c$ axis to one having its cavity axis perpendicular to the $c$ axis
would provide a relatively high sensitivity given by $\mathcal B_{1\Xi}^q$ and $\mathcal
B_{3\Xi}^q $. Another possibility would be two orthogonal cavities within a single block of
crystalline material. Such an arrangement might be favourable for eliminating parasitic effects,
like thermal expansion and vibration. However, in such an experiment the frequency change
$\delta_{\rm EM} \nu$ due to Lorentz violation in electrodynamics would not drop out, thus
complicating the analysis. Another configuration would be the comparison of a cavity against an
atomic clock, such as a cesium clock and/or a hydrogen maser. This scenario is interesting, since
it is projected for the OPTIS satellite \cite{OPTIS}. This sattelite is projected to carry three
cavities orthogonal to each other, so it is probably possible to separate the electrodynamic terms
from the electronic ones in an analysis of the complete data, that consists of all frequency
differences between the three cavities and the atomic clock(s). Such an analysis would also have
to take into account a possible Lorentz-violating shift of the atomic clock frequencies, which
could give the experiment sensitivity to additional parameters.


\section{Summary and Outlook}

We have calculated the change of the geometry of crystals that is caused by Lorentz invariance
violation in the fermionic sector of the extended standard model. The length change is caused by a
modified kinetic energy term $(\delta_{jk}+2E'_{jk})(1/2m) p_j p_k$  that enters the kinetic energy
term of the Hamiltonain for the free electron. $E'_{jk}=-c_{jk}-\frac 12 c_{00}
\delta_{jk}$, where $c_{\mu \nu}$ is a Lorentz tensor originating from the standard model
extension. The calculation proceeds using a Bloch ansatz for the wave function of the valence
electrons with the lattice periodic function given by a Fourier series. The crystal adjusts its
geometry such as to minimize its total energy. In that way, Lorentz violation in the electrons'
equation of motion affects the length of an electromagnetic cavity that is made from the crystal,
and thus the resonance frequency of a cavity made from the material. As a main result of this
paper, there is thus a method to measure the $c_{\mu \nu}$ in cavity tests of Lorentz violation.

Comparing cavities made from different materials, it is possible to separate the effect connected
to $c_{\mu \nu}$ from Lorentz violation in electrodynamics, that also affects the resonance
frequency of cavities. Under some assumptions that help to separate the electrodynamic and the
electronic terms, already performed experiments \cite{BrilletHall,MuellerASTROD,MuellerMM} imply
constraints on $c_{(XY)}$ and $c_{XX}-c_{YY}$ at the $10^{-14}$ level, and on
$c_{XX}+c_{YY}-2c_{ZZ}$ at the $10^{-12}$ level. This is to our knowledge the only present
experimental constraint on the components of $c_{\mu \nu}$. We discuss possible setups for
experiments that can obtain separate bounds without using these assumptions, and obtain results on
more components of $c_{\mu \nu}$. Future experiments on Earth and in space promise increased
sensitivity up to a part in $10^{18}$.

In the appendix, we briefly discuss the case of spin-polarized matter. An additional contribution
to the length change arises from a spin-dependent term $F'_{jkl} \frac{1}{m} p_j p_k \sigma^l$ also
originating from the standard model extension. This allows to deduce limits on $F'_{jkl}$ from
experiments using a spin-polarized cavity material, at least in principle.

Our model of the solid state could be improved by using material specific values for the Fourier
coefficients of the single electron wave function. Since our model Fourier coefficients already
satisfy the symmetry requirements for a realistic wave function, this might result in relatively
minor corrections for the length change. A check (maybe for a simple material), however, might be
worthwhile. Most importantly, however, a dedicated experiment will be performed to obtain more
complete and / or stronger limits on Lorentz violation in the electrons' equation of motion.
Cavities made from crystalline sapphire and fused quartz are ready to be implemented.

\acknowledgments We would like to thank V. Alan Kosteleck\'{y} for important comments. It is
a pleasure to acknowledge the cooperation with Stephan Schiller and to thank J\"{u}rgen Mlynek for
his valuable support.

\appendix

\section{Spin-polarized materials}\label{spinpolarization}

For a spin-polarized material, an additional contribution to the geometry change arises from a
spin-dependent term of the non relativistic single-electron hamiltonian of the SME, Eq.
({\ref{deltah}), that is given by $F'_{jkl}$. That means, from experiments using cavities made
from spin-polarized materials, a limit on $F'_{jkl}$ can, at least in principle, be deduced. This
is interesting, since many degrees of freedom of $F'_{jkl}$ are not yet fixed experimentally. In
this appendix, we estimate the effect and the level of sensitivity that can be expected for such
an experiment.

\subsection{Hamiltonian}
If the average of the spin expectation values is non-zero, the spin-dependent terms contribute to
the Lorentz-violating correction to the hamiltonian. We still assume a vanishing average helicity,
Eq. (\ref{helicity}), for all electrons. Thus,
\begin{eqnarray}\label{deltahspin}
\left<\delta h \right> & = & m c^2 B'_j \sum_{a=1}^N \left< (\sigma_a)^j \right>+ \frac{1}{m}
E'_{ij} \sum_{a=1}^N \left<(p_a)_i(p_a)_j\right> \nonumber \\ & & + \frac 1m F'_{jkl} \sum_{a=1}^N
\left< (p_a)_j(p_a)_k (\sigma_a)^l \right> \, .
\end{eqnarray}
We assume that a fraction $\eta^i$ of the total $N$ electrons have their spin $\frac 12 $
polarized parallel to $x^i$. The other electrons are assumed to be unpolarized. The average of the
spin expectation values is then given by
\begin{equation}
S^l \equiv \frac 1N \sum_a \left< (\sigma_a)^l \right>  =  \frac 12 \eta^l \, .
\end{equation}
For the last term of Eq. (\ref{deltahspin}),
\begin{equation}
\sum_{a=1}^N \left< (p_a)_j(p_a)_k (\sigma_a)^l \right> =  \frac 12 \eta^l \sum_{a=1}^N \left<
(p_a)_j(p_a)_k \right> \, .
\end{equation}
Therefore,
\begin{eqnarray}
\left<\delta h \right> & = & m c^2 \frac N2 B'_j \eta^j \\ & & + \frac{1}{m} \left( E'_{ij} +
\frac 12 F'_{ijk} \eta^k \right) \sum_{a=1}^N \left<(p_a)_i(p_a)_j\right>\, . \nonumber
\end{eqnarray}

\subsection{Geometry change}

The term $ m c^2 N B'_j \eta^j/2$ contained in the hamiltonian is independent from the crystal
geometry and does, therefore, not lead to a geometry change. The second term that is proportional
to the average of $\left< p_i p_j \right>$ over all electrons, however, leads to a geometry
change, that can be calculated in analogy to the discussion in the main parts of this paper. We
can overtake the result, Eq. (\ref{edctensor}), for the geometry change if we replace $E'_{ij}$ by
$\tilde E'_{ij} = E'_{ij} + \frac 12 F'_{ijk}\eta^k$.

The sensitivity of the cavity geometry to $F'_{jkl}$ is thus given by $\mathcal B_{\Gamma \Xi}$ as
well as $\eta^k$. The magnitude of the latter can be estimated as the ratio of the number of
spin-polarized electrons $n_{B,u}$ to the total number of electrons $N_{e,u}$ per unit cell,
$|\eta|=n_{B,u}/n_{e,u}$. In a saturated ferromagnetic material, e.g. iron at a magnetic field of
$1.7$\,T, $n_{B,u}\approx 2.2$ \cite{Kittel} spins are polarized per unit cell, so $|\eta|$ is of
order unity. (Note that $n_{B,u}$ can be as high as $\simeq 10$ for Dysprosium.) Therefore, the
sensitivity of the cavity length (and thus, its resonance frequency) to $F'_{jkl}$ is comparable
to the sensitivity to $E'_{jk}$.

\subsection{Possible experiments}

If the cavity is made from a spin-polarized solid, i.e., a magnetized ferromagnetic material, the
cavity length would depend on $E'_{ij}+ F'_{ijk} \eta^k/2$. That means, from a measurement of the
resonance frequency of such a cavity, a limit on $F'_{ijk}$ could be derived, provided that
separate limits on $E'_{ij}$ are known from previous experiments using one of the methods
discussed above. However, note that the systematics of such an experiment are largely unknown. The
selection of materials suitable for building stable cavities is a highly nontrivial discussion of
the experimental systematics, some of which are far from obvious. Because of the manifold effects
connected to magnetization (e.g., magnetostriction), an experiment using a magnetized cavity could
suffer from various systematic effects, so our discussion is a bit speculative. A theoretical
complication is that practical ferromagnetic materials are usually alloys (such as AlNiCo),
whereas the theory presented above is directly applicable for crystals only.

For such a measurement of $F'_{ijk}$, one could use a cavity made from a permanent magnetic
material. The direction of the spin polarization with respect to the cavity axis determines the
components of $F'_{ijk}$ which dominate the experiment. A rotation of the cavity would modulate
the $F'_{ijk}$-induced frequency shift. The corresponding time dependency of the cavity resonance
frequency would be the signal for a non-zero $F'_{ijk}$. It could be beneficial to use solely the
Earth's rotation to avoid possible systematics associated with a magnetized cavity rotating in the
Earth's magnetic field. Magnetic shielding will probably also be necessary. If the frequency
stability of a cavity made from a suitable magnetized material would be of the same order as the
stability achieved with quartz or sapphire cavities, limits on $F'_{ijk}$ of order $10^{-15}$
could be achieved.

\section{Notation convention in elasticity theory}\label{notationconvention}

For obtaining the material specific values of the sensitivity tensor $\mathcal B_{abcd}$, the
compliance tensor $\mu_{abcd}$ has to be known. The relation between stress $\sigma_{ij}$ and
strain $e_{dc}$ is given as
\begin{equation}
\sigma_{ij} = \lambda_{ijkl} e_{kl} \, .
\end{equation}
In engineering, it is common to replace this by a six-dimensional matrix equation
\begin{equation}
{\bf \sigma} = {\bf S} \cdot {\bf e} \, ,
\end{equation}
where \cite{Ashcroft}, p.445
\begin{eqnarray}
{\bf \sigma} & = & (\sigma_{xx}, \sigma_{yy}, \sigma_{zz}, \sigma_{yz}, \sigma_{zx}, \sigma_{xy})
\, ,\label{stresssixvector} \\ {\bf e} & = & (e_{xx}, e_{yy}, e_{zz}, 2e_{yz}, 2e_{zx}, 2e_{xy})
\, ,\label{strainsixvector}
\end{eqnarray}
and
\begin{equation}
{\bf S} = \left( \begin{array}{cccccc}
  \lambda_{1111} & \lambda_{1122} & \lambda_{1133} & \lambda_{1123} & \lambda_{1131} & \lambda_{1112} \\
  \lambda_{2211} & \lambda_{2222} & \lambda_{2233} & \lambda_{2223} & \lambda_{2231} & \lambda_{2212} \\
  \lambda_{3311} & \lambda_{3322} & \lambda_{3333} & \lambda_{3323} & \lambda_{3331} & \lambda_{3312} \\
  \lambda_{3211} & \lambda_{3222} & \lambda_{3233} & \lambda_{3223} & \lambda_{3231} & \lambda_{3212} \\
  \lambda_{3111} & \lambda_{3122} & \lambda_{3133} & \lambda_{3123} & \lambda_{3131} & \lambda_{3112} \\
  \lambda_{1211} & \lambda_{1222} & \lambda_{1233} & \lambda_{1223} & \lambda_{1231} & \lambda_{1212} \\
\end{array} \right) \, .
\end{equation}
This is called the Voigt convention in the literature \cite{JonesMarch}, pp. 604-609. The matrix
${\bf S}$ is symmetric, it thus contains at most 21 independent elements. The symmetry of the
crystal reduces the number of independent elements. For example, the matrix for cubic symmetry has
three independent elements:
\begin{equation}
{\bf S} = \left( \begin{array}{cccccc} S_{11} & S_{12} & S_{12} & 0 & 0 & 0 \\ S_{12} & S_{11} &
S_{12} & 0 & 0 & 0 \\ S_{12} & S_{12} & S_{11} & 0 & 0 & 0 \\ 0 & 0 & 0 & S_{44} & 0 & 0 \\ 0 & 0
& 0 & 0 & S_{44} & 0 \\ 0 & 0 & 0 & 0 & 0 & S_{44} \end{array} \right) \, .
\end{equation}
For trigonal symmetry, there are six:
\begin{equation}
{\bf S} = \left( \begin{array}{cccccc} S_{11} & S_{12} & S_{13} & S_{14} & 0 & 0 \\ S_{12} &
S_{11} & S_{13} & -S_{14} & 0 & 0 \\ S_{13} & S_{13} & S_{33} & 0 & 0 & 0 \\ S_{14} & -S_{14} & 0
& S_{44} & 0 & 0
\\ 0 & 0 & 0 & 0 & S_{44} & S_{14} \\ 0 & 0 & 0 & 0 & S_{14} & S_{66}
\end{array}
\right)
\end{equation}
where
\begin{equation}
S_{66}= 2(S_{11}-S_{12}) \, .
\end{equation}
By inverting ${\bf S}$ one obtains the compliance matrix ${\bf C}$ that satisfies ${\bf S}\cdot
{\bf C}=\openone$, where $\openone$ is the six dimensional unit matrix. ${\bf C}$ enters the
equation ${\bf e}={\bf C} \cdot {\bf \sigma}$ between the stress and strain 6-vectors Eqs.
(\ref{stresssixvector},\ref{strainsixvector}). On the other hand, the compliance {\em tensor}
$\mu_{abcd}$ enters the relationship between the stress and strain {\em tensors}, $e_{ab}
=\mu_{abcd} \sigma_{cd}$. Therefore, the compliance tensor is related to the compliance matrix by
\begin{equation}\label{compliancematrix}
{\bf C}=\left( \begin{array}{cccccc}
  \mu_{1111} & \mu_{1122} & \mu_{1133} & 2\mu_{1123} & 2\mu_{1131} & 2\mu_{1112} \\
  \mu_{2211} & \mu_{2222} & \mu_{2233} & 2\mu_{2223} & 2\mu_{2231} & 2\mu_{2212} \\
  \mu_{3311} & \mu_{3322} & \mu_{3333} & 2\mu_{3323} & 2\mu_{3331} & 2\mu_{3312} \\
  2\mu_{3211} & 2\mu_{3222} & 2\mu_{3233} & 4\mu_{3223} & 4\mu_{3231} & 4\mu_{3212} \\
  2\mu_{3111} & 2\mu_{3122} & 2\mu_{3133} & 4\mu_{3123} & 4\mu_{3131} & 4\mu_{3112} \\
  2\mu_{1211} & 2\mu_{1222} & 2\mu_{1233} & 4\mu_{1223} & 4\mu_{1231} & 4\mu_{1212} \\
\end{array} \right) \, .
\end{equation}
The compliance and elasticity matrices for cubical crystals have the same symmetry; for trigonal
crystals, the symmetry of the compliance matrix is similar to the one of the elasticity matrix,
with the exception that $C_{66}= \frac 12 (C_{11}-C_{12})$. The compliance tensor elements
$\mu_{abcd}$ are obtained from the tabulated elements of the elasticity matrix ${\bf S}$ by
inverting the elasticity matrix and reading of the tensor elements from Eq.
(\ref{compliancematrix}).

\section{Signal Components for laboratory Experiments with turntable}\label{sigcomp}

Here, we give the full signal components caused by Lorentz violation in the electrons' equation of
motion in the laboratory frame for a cavity rotated, using a turntable, at an angular frequency
$\ot$, assuming a material of trigonal or higher crystal symmetry, i.e., the sensitivity matrix
$\mathcal B$ is of the form Eq. (\ref{matrixtrigonal}) or simpler. The rotation axis is fixed to
point vertically. We use a turntable time scale $t_t$ defined such that $t_t = 0$ at any one
instant when the cavity is pointing in the $x$ direction of the laboratory frame.

We use two reference frames, one sun-centered celestial equatorial reference frame and one
laboratory frame. As defined in \cite{Kostelecky99}, the sun-centered frame has the $X$ axis
pointing towards the vernal equinox (spring point) at 0\,h right ascension and 0$^\circ$
declination, the $Z$ axis pointing towards the celestial north pole ($90^\circ$ declination) and
the $Y$ axis such as to complete the right handed orthogonal dreibein. Earth's equatorial plane
lies in the $X-Y$ plane; its orbital plane is tilted by $\eta \simeq 23^\circ$ with respect to the
latter. The time scale $T=0$ when the sun passes the spring point, e.g., on march 20, 2001 at
13:31 UT. \\ The laboratory frame has the $x$ axis pointing south, the $y$ axis east, and the $z$
axis vertically upwards. The laboratory time scale $\To = 0$ when the $y$ and the $Y$ axis
coincide.

The signal derivation starts from the symmetrized tensor $c_{(\mu \nu)}$ given in the sun-centered
celestial equatorial reference frame, which is suitable for expressing the tensor because it is
inertial on all time-scales involved in terrestrial experiments. To $c_{(\mu \nu)}$, we first
apply a Lorentz boost to first order in $\beta_\oplus \simeq 10^{-4}$, the velocity of Earth's
orbit
\begin{equation} \vec \beta_{\oplus} = \beta_\oplus \left(\begin{array}{c} \sin \Oo T \\
-\cos \eta \cos \Oo T
\\ -\sin \eta \cos \Oo T \end{array} \right) \, ,
\end{equation}
where $\Oo \simeq 2\pi/1$\,yr is the angular frequency of Earth's orbit. We neglect the smaller
velocity $0 < \beta_L \lesssim 1.5\times 10^{-6}$ due to Earth's rotation in order not to
complicate the signal components below further. Subsequently, application of the rotation matrix
\begin{equation}\label{rotationmatrix}
R = \left( \begin{array}{ccc}
   \cos \chi \cos \oo \To & \cos \chi \sin \oo \To & -\sin \chi \\
   -\sin \oo \To & \cos \oo \To  & 0 \\
   \sin \chi \cos \oo \To & \sin \chi \sin \oo \To & \cos \chi
\end{array} \right) \, ,
\end{equation}
where $\chi$ is the geographical colatitude, and $\oo \simeq 2\pi/$23\,h\,56\,min Earth's rotation
angular frequency, leads to the tensor $c_{\mu\nu}$ as expressed within the laboratory frame.
Another rotation around the $z$ axis using the rotation matrix
\begin{equation}
R_t = \left( \begin{array}{ccc} \cos \ot t_t & \sin \ot t_t & 0 \\  -\sin \ot t_t & \cos \ot t_t &
0
\\ 0 & 0 & 1 \end{array} \right)
\end{equation}
leads to the quantities within the rotating turntable frame, which are then decomposed according
to Eq. (\ref{hamiltonianparameters}). The time scale $t_t=0$ when the cavity axis is parallel to
the $x$ axis. Insertion of the results into Eq. (\ref{lchange6times6}) gives the cavity length
change, and thus the frequency change that is given below.

For compact notation, we define the abbreviations
\begin{eqnarray}
\omega(a,b,c) & = & a \ot + b \oo + c \Oo \, , \\ \phi(a,b,c) & = & a \ot t_t + b \oo \To + c \Oo
T \, .
\end{eqnarray}
We give the signal components for the two most interesting cavity constructions: A cavity with the
resonator axis pointing parallel to the crystals's $c$ axis (which is currently the most familiar
cavity type), and a cavity with the resonator axis parallel to the crystal's $a$ or $b$ axis
(which gives sensitivity to the $c_{(0i)}$ components to first order in $\beta_\oplus$, the
Earth's orbital velocity). The signals are expressed as a Fourier series
\begin{eqnarray}
\frac{\delta \nu}{\nu} & = & C(0,0,0)+ \sum_{a,b,c} [ S(a,b,c) \sin \phi(a,b,c) \\ & & + C(a,b,c)
\cos \phi(a,b,c)] \nonumber
\end{eqnarray}
with coefficients $S(a,b,c)$ and $C(a,b,c)$; the dc component $C(0,0,0)$ is not included in the
equations below, as it is not measurable.

\subsection{Cavity axis parallel to $a$ axis}
We use the abbreviations $\mathcal B_A = \mathcal B_{11} + \mathcal B_{12} - 2 \mathcal B_{13}$
and $\mathcal B_B = \mathcal B_{11}-\mathcal B_{12}$. The signal consists of 18 frequencies
$\omega(a,b,c)$ with
\begin{eqnarray*}
C(0,1,0) & = & -\mathcal B_A c_{(XZ)} \cos \chi \sin \chi \, , \\ S(0,1,0) & = & -\mathcal B_A
c_{(YZ)} \cos \chi \sin \chi \, , \\ C(0,2,0) & = & -(1/4) \mathcal B_A(c_{XX}-c_{YY})\sin^2\chi
\, , \\ S(0,2,0) & = & -(1/2)\mathcal B_Ac_{(XY)}\sin^2\chi \, , \\
 C(1,-2,0) & = & 2\mathcal B_{14}c_{(XY)} \cos \frac \chi 2
\sin^3\frac\chi2 \, , \\ S(1,-2,0) & = & \mathcal
B_{14}(c_{XX}-c_{YY})\cos\frac\chi2\sin^3\frac\chi2 \, ,
\\ C(1,-1,-1) & = & -\mathcal B_{14}\beta_\oplus
\sin^2\frac\chi2\sin\frac\eta2\left(c_{(TY)}\cos\frac\eta2 \right. \\ & & \left.
+c_{(TZ)}\sin\frac\eta2\right) \, , \\ S(1,-1,-1) & = & -\frac12\mathcal B_{14}\beta_\oplus
c_{(TX)}\sin^2\frac\chi2\sin\eta \, , \\ C(1,-1,0) & = & \mathcal
B_{14}c_{(YZ)}(1+2\cos\chi)\sin^2\frac\chi2 \, , \\ S(1,-1,0) & = & \mathcal B_{14} c_{(XZ)}
(1+2\cos\chi)\sin^2\frac\chi2 \, , \\ C(1,-1,1) & = & \mathcal
B_{14}\beta_\oplus\cos\frac\eta2\sin^2\frac\chi2\left(c_{(TZ)}\cos\frac\eta2 \right. \\ & & \left.
-c_{(TY)}\sin\frac\eta2\right) \, , \\  S(1,-1,1) & = & -\frac12\mathcal B_{14}\beta_\oplus
c_{(TX)}\sin^2\frac\chi2\sin\eta \, , \\ C(1,0,-1) & = & -(1/2)\mathcal B_{14}\beta_\oplus
c_{(TX)}\cos\eta\sin\chi \, , \\ S(1,0,-1) & = & (1/2)\mathcal B_{14}\beta_\oplus c_{(TY)}\sin\chi
\, , \\ C(1,0,0) & = & 0 \, , \\ S(1,0,0) & = & -(1/4)\mathcal B_{14}(c_{XX}+c_{YY}-2c_{ZZ})\sin 2
\chi \, , \\ C(1,0,1) & = & -(1/2)\mathcal B_{14}\beta_\oplus c_{(TX)}\cos\eta\sin\chi \, , \\
S(1,0,1) & = & -(1/2)\mathcal B_{14}\beta_\oplus c_{(TY)}\sin\chi \, ,
\\ C(1,1,-1) & = & -\mathcal B_{14}\beta_\oplus\cos^2\frac\chi2\cos\frac\eta2\left(c_{(TZ)}\cos\frac\eta2 \right. \\ & & \left.
-c_{(TY)}\sin\frac\eta2\right) \, , \\ S(1,1,-1) & = & -\frac12\mathcal B_{14}\beta_\oplus
c_{(TX)}\cos^2\frac\chi2\sin\eta \, , \\ C(1,1,0) & = & \mathcal B_{14}c_{(YZ)} \cos^2\frac \chi2
(2\cos \chi-1) \, , \\ S(1,1,0) & = & -\mathcal B_{14}c_{(XZ)}\cos^2\frac\chi2(2\cos\chi-1) \, ,
\\ C(1,1,1) & = & \mathcal B_{14} \beta_\oplus\cos^2\frac\chi2\sin\frac\eta2\left(c_{(TY)}\cos\frac\eta2
\right. \\ & & \left. +c_{(TZ)}\sin\frac\eta2\right) \, , \\ S(1,1,1) & = & -\frac12\mathcal
B_{14}\beta_\oplus c_{(TX)}\cos^2\frac\chi2\sin\eta \, , \\ C(1,2,0) & = & (1/2)\mathcal
B_{14}c_{(XY)}(1+\cos\chi)\sin\chi \, , \\ S(1,2,0) & = & -\mathcal
B_{14}(c_{XX}-c_{YY})\cos^3\frac\chi2\sin\frac\chi2 \, , \\ C(2,-2,0) & = & \frac12\mathcal
B_B(c_{XX}-c_{YY})\sin^4\frac\chi2 \, , \\ S(2,-2,0) & = & -\mathcal B_Bc_{(XY)} \sin^4\frac\chi2
\, , \\ C(2,-1,0) & = & 2\mathcal B_B c_{(XZ)} \cos \frac\chi2\sin^3\frac\chi2 \, , \\ S(2,-1,0) &
= & -2\mathcal B_Bc_{(YZ)}\cos\frac\chi2\sin^3\frac\chi2 \, , \\ C(2,0,0) & = & -(1/4) \mathcal
B_B (c_{XX}+c_{YY}-2c_{ZZ})\sin^2\chi \, , \\ S(2,0,0) & = & 0 \, , \\ C(2,1,0) & = & -2\mathcal
B_Bc_{(XZ)} \cos^3\frac\chi2\sin\frac\chi2 \, , \\ S(2,1,0) & = & -2\mathcal
B_Bc_{(YZ)}\cos^3\frac\chi2\sin\frac\chi2 \, , \\ C(2,2,0) & = & \frac12\mathcal
B_B(c_{XX}-c_{YY})\cos^4\frac\chi2 \, , \\ S(2,2,0) & = & \mathcal B_Bc_{(XY)}\cos^4\frac\chi2 \,
.
\\
\end{eqnarray*}
The signal for a cavity parallel to the crystals $b$ axis can be obtained from these equations, if
the $x$ and $y$ axis are interchanged.

\subsection{Signal for a cavity parallel to the $c$ axis}
The cavity is oriented with its axis parallel to the $x_t$ axis. We introduce the abbreviations
$\mathcal B_C = -2\mathcal B_{31}+\mathcal B_{32}+\mathcal B_{33}$ and $\mathcal B_D = \mathcal
B_{32}-\mathcal B_{33}$. (For the trigonal case, Eq. (\ref{matrixtrigonal}), $\mathcal
B_{32}=\mathcal B_{31}/3$, for isotropic materials, $\mathcal B_{32}=\mathcal B_{31}$.) We have
seven signal frequencies with the amplitudes
\begin{eqnarray*}
C(0,1,0) & = & -\mathcal B_C c_{(XZ)}\cos\chi\sin\chi \, , \\ S(0,1,0) & = & -\mathcal B_C
c_{(YZ)} \cos \chi \sin \chi \, , \\ C(0,2,0) & = & -(1/4)\mathcal B_C(c_{XX}-c_{YY})\sin^2\chi \,
, \\ S(0,2,0) & = &- (1/2)\mathcal B_C c_{(XY)} \sin^2\chi \, , \\ C(2,-2,0) & = &-
\frac12\mathcal B_D(c_{XX}-c_{YY})\sin^4\frac\chi2 \, , \\ S(2,-2,0) & = & \mathcal
B_Dc_{(XY)}\sin^4\frac\chi2 \, ,
\\ C(2,-1,0) & = & -2\mathcal B_D c_{(XZ)}\cos \frac\chi2 \sin^3\frac\chi2 \, , \\ S(2,-1,0) & = &
2\mathcal B_Dc_{(YZ)}\cos\frac\chi2\sin^3\frac\chi2 \, , \\ C(2,0,0) & = & (1/4)\mathcal
B_D(c_{XX}+c_{YY}-2c_{ZZ})\sin^2\chi \, , \\ S(2,0,0) & = & 0 \, ,
\\C(2,1,0) & = & 2\mathcal B_Dc_{(XZ)}\cos^3\frac\chi2\sin\frac\chi2 \, , \\ S(2,1,0) & = & 2\mathcal B_Dc_{(YZ)}
\cos^3\frac\chi2 \sin\frac \chi2 \, , \\ C(2,2,0) & = & -\frac12\mathcal
B_D(c_{XX}-c_{YY})\cos^4\frac\chi2 \, , \\ S(2,2,0) & = & -\mathcal B_Dc_{(XY)}\cos^4\frac\chi2 \,
.
\end{eqnarray*}



\begin{references}
\bibitem{noinversionsymmetry} Without using Eq. (\ref{cminusn}) (for which we assumed a crystal
with inversion symmetry),  these terms can still be eliminated in the crystal's rest frame by
using the vanishing of the average momentum expectation value. $\sum_a \left< (p_a)_i \right>=0$
implies
\[
0  =  \sum_a\sum_{\vec n}|(c_{\vec n})_a|^2[(q_a)_i + \hbar n_jk_{ij}] \, ,
\]
which is derived in analogy to Eq. (\ref{momentumsquared}). The quasi-momenta $q_a$ are within the
first Brillouin zone. Thus, as explained after Eq. (\ref{dxeinsintegral}), the two terms cannot
cancel each other, so
\[
\sum_a \sum_{\vec n} |(c_{\vec n})_a|^2 n_jk_{ij}=0 \, .
\]
\end{references}
\end{document}